\documentclass[prd,preprint,superscriptaddress,nofootinbib]{revtex4}
\usepackage[utf8]{inputenc}
\usepackage{amsmath,amssymb}
\usepackage{graphicx}
\usepackage{color,slashed}
\pdfoutput=1

\begin{document}
\title{\bf Muon $g-2$ in a two-Higgs-doublet model with a type-II seesaw mechanism}

\author{Chuan-Hung Chen}
\email[e-mail: ]{physchen@mail.ncku.edu.tw}
\affiliation{Department of Physics, National Cheng-Kung University, Tainan 70101, Taiwan}
\affiliation{Physics Division, National Center for Theoretical Sciences, Taipei 10617, Taiwan}

\author{Cheng-Wei Chiang}
\email[e-mail: ]{chengwei@phys.ntu.edu.tw}
\affiliation{Department of Physics and Center for Theoretical Physics, National Taiwan University, Taipei 10617, Taiwan}
\affiliation{Physics Division, National Center for Theoretical Sciences, Taipei 10617, Taiwan}

\author{Takaaki Nomura}
\email{nomura@scu.edu.cn}
\affiliation{College of Physics, Sichuan University, Chengdu 610065, China}

\date{\today}

\begin{abstract}
We study the two-Higgs-doublet model with type-II seesaw mechanism.  In view of constraints from the Higgs data, we consider the aligned two-Higgs-doublet scheme and its effects on muon anomalous magnetic dipole moment, $a_{\mu}$, including both one-loop and two-loop Barr-Zee type diagrams.  Thanks to a sizable trilinear scalar coupling, the Barr-Zee type diagrams mediated by the Higgs triplet fields have a dominant effect on $a_{\mu}$.  In particular, unlike the usual two-Higgs-doublet models that require exotic Higgs bosons light in mass, the masses of the corresponding particles in the model are of ${\cal O}(100)$~GeV.  The doubly-charged Higgs boson presents a different decay pattern from the usual Higgs triplet model and thus calls for a new collider search strategy, such as multi-$\tau$ searches at the LHC.
\end{abstract}
\maketitle

\section{Introduction}

A long-standing anomaly in particle physics is the muon anomalous magnetic dipole moment (dubbed the muon $g-2$ anomaly) denoted by $a_\mu \equiv (g-2)_\mu/2$, where the data and Standard Model (SM) show an over $3\sigma$ disagreement.  The E821 experiment at Brookhaven National Lab (BNL) has presented a precision measurement of:
 \begin{equation}
 a^{\rm exp}_\mu=11 659 2089(63)\times 10^{-11} \,,
 \end{equation}
 with an uncertainty of $0.54$~ppm~\cite{Bennett:2006fi}.
The current theoretical estimate of $a_\mu$ within the SM has also reached a comparable precision of $0.369$~ppm, and is shown to be~\cite{Aoyama:2020ynm}:
 \begin{align}
 a^{\rm SM}_\mu = 116591810(43) \times 10^{-11}\,. 
 \end{align}
The deviation between the experiment and the SM is $\Delta a_\mu = a^{\rm exp}_\mu -a^{\rm SM}_\mu= 279(76)\times 10^{-11}$ with an achievement of $3.7\sigma$. 
The new muon $g-2$ measurement performed in the E989 Run 1 experiment at Fermilab, designed to have a precision of $0.14$~ppm, reports its first measurement as~\cite{Abi:2021gix}:
 \begin{equation}
  a^{\rm FNAL}_\mu=116592040(54)\times 10^{-11} \,.
 \end{equation}
Combining all available measurements on the quantity, we now have a $4.2\sigma$ deviation between experiment and SM expectation\footnote{The latest lattice QCD calculation for the leading hadronic vacuum polarization from the BMW collaboration is obtained as $a^{\rm LO-HVP}_\mu=707.5(5.5)\times 10^{-10}$, which leads to a larger $a_\mu$, can be found in~\cite{Borsanyi:2020mff}.}, accentuating the muon $g-2$ anomaly.

On the other hand, since the discovery of Higgs boson at the LHC in 2012 summer, measurements of the Higgs signal strengths, commonly used as a measure of deviations from the SM, have been improving over the years.  They are found to be quite consistent with the SM expectations and, hence, models with extensions in the scalar sector are severely constrained.  One possibility for a new physics (NP) model to achieve such a good agreement with the SM in the Higgs couplings while having exotic Higgs bosons of mass at ${\cal O}(100)$~GeV scale is when the model shows the so-called alignment limit~~\cite{Gunion:2002zf,Carena:2013ooa,Bernon:2015qea}.

In this work, we study the contributions of a model with an extended scalar sector to the muon $g-2$ when the relevant theoretical and experimental constraints are taken into account.  One purpose is to revisit the two-Higgs-doublet models (2HDMs), where the earlier studies can be found in Refs.~\cite{Leveille:1977rc,Haber:1978jt,Krawczyk:1996sm,Dedes:2001nx,Cheung:2001hz,Larios:2001ma,Chen:2001kn,Arhrib:2001xx,Krawczyk:2002df,Gunion:2008dg,Broggio:2014mna,Ilisie:2015tra,Han:2015yys,Li:2018aov,Rose:2020nxm,Sabatta:2019nfg,Jana:2020pxx,Li:2020dbg,Chen:2021rnl,Han:2018znu,Botella:2020xzf}.  It is known that to explain the muon $g-2$ in this framework, the new scalar or pseudoscalar boson are required to be as light as ${\cal O}(10)$~GeV.  Although such a parameter space is still allowed by the current data, it is of interest to probe the scenarios where the new scalar masses can be more relaxed and of $\sim {\cal O}(100)$~GeV by further extending the scalar sector.  More importantly, such a new extension should also address some other unsolved issues, such as the origin of neutrino mass, that the simple 2HDMs cannot accommodate.

To achieve the above-mentioned goals, we consider the 2HDM with type-II seesaw mechanism~\cite{Chen:2014xva,Ouazghour:2018mld}.  In addition to the SM Higgs doublet, the scalar sector contains another complex doublet and a complex triplet.  Moreover, we will consider the so-called aligned two-Higgs-doublet scheme (A2HDS), where the Yukawa couplings of the two Higgs doublets to the SM fermions are proportional to each other and one of the neutral physical Higgs boson is SM-like.  The A2HDS has the interesting feature that it reduces to various 2HDM types by taking proper limits on the alignment parameters.  With a small vacuum expectation value (VEV) induced by electroweak symmetry breaking from the two Higgs doublets, the Higgs triplet in the model provides Majorana mass to neutrinos through the so-called type-II seesaw mechanism~\cite{Magg:1980ut,Lazarides:1980nt,Schechter:1980gr,Cheng:1980qt,Mohapatra:1980yp,Bilenky:1980cx}.

It is found that rather than a simple combination of the 2HDM and the type-II seesaw model (also called the Higgs triplet model or HTM), the model presents several interesting features:
\begin{enumerate}
\item 
The coupling between the heavier neutral Higgs boson in the 2HDM and the doubly-charged Higgs boson in the HTM can significantly enhance the muon $g-2$ through two-loop Barr-Zee type diagrams~\cite{Barr:1990vd,Chang:2000ii}, even when the heavier neutral Higgs mass is $\sim {\cal O}(100)$~GeV.

\item
The Higgs triplet VEV is now determined by three lepton number-violating parameters instead of just one in the simple HTM.  As a result of the extra freedom, these parameters are not necessarily of the same order as the Higgs triplet VEV~\cite{Chen:2014xva}.

\item
With a sizable Higgs triplet VEV, the doubly-charged Higgs boson shows a richer decay pattern.  As a result, the doubly-charged Higgs boson can evade the recent ATLAS lower bound of 350~GeV in pair production~\cite{Aad:2021lzu}. In addition to the like-sign diboson channel, the doubly-charged Higgs boson can also be probed via channels involving the light charged Higgs boson. 
\end{enumerate}

The paper is organized as follows.  In Sec.~\ref{sec:model}, we derive the Yukawa couplings in the A2HDS and show the relations between the scheme and the various types of 2HDMs with $\mathbb{Z}_2$ symmetry.  The mass-square relations of the triplet Higgs bosons are discussed, and the CP-even neutral Higgs couplings with the charged Higgses are given.  In Sec.~\ref{sec:loop_g2}, we discuss the results of one-loop and the dominant two-loop Barr-Zee type diagrams. Using the bounded parameters, we present the detailed numerical analysis and discussion in Sec.~\ref{sec:NA}.  Sec.~\ref{sec:summary} summarizes our findings in this work.  The full scalar mass matrices and their approximations in the limit of neglecting $v_\Delta$ are given in appendix~\ref{app:mass_matrix}.

\section{Model and Interactions}\label{sec:model}

We consider a model where the scalar sector is extended with a doublet with $Y = 1/2$ and a complex triplet with $Y = 1$.  In the following, we discuss the general Yukawa interactions and scalar potential in this model.

\subsection{Scalar potential and the trilinear  scalar couplings
\label{sec:Scalar potential}}

Since the scalar sector is an extension of 2HDM or of type-II seesaw, in the following, we briefly discuss the essential parts for our analysis. 
First, as we will assume negligibly small mixing between the doublet fields and the triplet field, it is useful to go to the Higgs basis in the usual 2HDM, defined by: 
 \begin{equation}
\begin{pmatrix}
             H_1 \\
              H_2\\
\end{pmatrix}
 = 
\begin{pmatrix}
  c_\beta &  s_\beta &   \\
 -s_\beta  &  c_\beta  &  
\end{pmatrix}
\begin{pmatrix}
             \Phi_1 \\
              \Phi_2\\
\end{pmatrix}
\,, \label{eq:Higgs_Basis}
 \end{equation}
where $v_i$ is the VEV of $\Phi_i$ ($i = 1,2$), $c_\beta~(s_\beta) = \cos\beta~ (\sin \beta)$,  $\tan\beta = v_2/v_1$ and $v=\sqrt{v^2_1+ v^2_2} \simeq 246$~GeV.  Written in terms of field components, the Higgs doublets $H_{1,2}$ and triplet $\Delta$ are:
\begin{align}
H_1&= \begin{pmatrix}
              G^+ \\
              (v+H^0_1 +i G^0)/\sqrt{2} \\
\end{pmatrix}
\,,~~ 
H_2= \begin{pmatrix}
              H^+ \\
              (H^0_2 +i A^0)/\sqrt{2} \\
\end{pmatrix}
\,, \nonumber \\
\Delta & = \begin{pmatrix}
    \delta^+/\sqrt{2} & \delta^{++}  \\ 
    (v_\Delta + \delta^0 + i\eta^0)/\sqrt{2} & ~ -\delta^+/\sqrt{2} \\ 
\end{pmatrix}
\,. \label{eq:scalars}
\end{align}
In the conventional CP-conserving 2HDM, $G^{\pm(0)}$ are the Goldstone bosons, and $H^\pm$ and $A^0$ are the charged Higgs boson and the CP-odd pseudoscalar physical states.  In addition, the CP-even scalars $H^0_1$ and $H^0_2$ mix to give their mass eigenstates via: 
 \begin{equation}
\begin{pmatrix}
             H \\
              h\\
\end{pmatrix}
 = 
\begin{pmatrix}
 c_{\beta-\alpha} &  - s_{\beta-\alpha}  \\
 s_{\beta-\alpha}  &  c_{\beta-\alpha}  
\end{pmatrix}
\begin{pmatrix}
             H^0_1 \\
              H^0_2\\
\end{pmatrix}
\,, 
 \label{eq:2HDM_mass_Basis}
 \end{equation} 
where $h$ is the 125-GeV SM-like Higgs boson, $c_{\beta-\alpha}=\cos(\beta-\alpha)$, $s_{\beta-\alpha}=\sin(\beta-\alpha)$, and $\alpha$ is the mixing angle of $\Phi^0_1$ and $\Phi^0_2$. Although $\delta^0$, $\eta^0$ and $\delta^\pm$ generally mix with $(H, h)$, $(G^0, A^0)$, and $(G^\pm, H^\pm)$, respectively, such mixings are small and phenomenologically negligible when $v_\Delta\ll 1$~GeV, as is the case considered in this work.  Hence, it is a good approximation to take $h$, $H$, $A^0$, and $H^\pm$ as the physical states.

The scalar potential of two-Higgs-doublet fields and the Higgs triplet field under the $SU(2)_L\times U(1)_Y$ gauge symmetry is given by:
 \begin{equation}
 V= V(\Phi_1, \Phi_2)+ V(\Delta) + V(\Phi_1,\Phi_2,\Delta)\,, 
 \end{equation}
where each term is more explicitly given by~\cite{Chen:2014xva,Ouazghour:2018mld}
  \begin{align}
V(\Phi_1,\Phi_2) 
=& m^2_1 \Phi^\dagger_1 \Phi_1 + m^2_2 \Phi^\dagger_2 \Phi_2 - m^2_{12} ( \Phi^\dagger_1 \Phi_2 + \mbox{H.c.})
+ \frac{1}{2} \lambda_1 ( \Phi^\dagger_1 \Phi_1)^2   \nonumber \\
 &+  \frac{1}{2} \lambda_2 (\Phi^\dagger_2 \Phi_2)^2 + \lambda_3  \Phi^\dagger_1 \Phi_1 \Phi^\dagger_2 \Phi_2 + \lambda_4  \Phi^\dagger_1 \Phi_2 \Phi^\dagger_2 \Phi_1 +  \left[\frac{1}{2}\lambda_5(\Phi^\dagger_1 \Phi_2)^2  \right. \nonumber \\
 &\left.+ \lambda_6 (\Phi^\dag_1 \Phi_1) (\Phi^\dag_{2} \Phi_1) + \lambda_7 (\Phi^\dag_2 \Phi_2) (\Phi^\dag_2 \Phi_1)
 + \mbox{H.c.} \right] \,,  \label{eq:v2_a}
\\
 %
V(\Delta )
=& m^2_\Delta Tr \Delta^\dagger \Delta + \lambda_{\Delta 1} (Tr \Delta^\dagger \Delta)^2 + \lambda_{\Delta 2} Tr (\Delta^\dagger \Delta)^2\,,  \label{eq:v2_b}
\\
 %
V(\Phi_1,\Phi_2,\Delta) 
=& \left( \mu_1 \Phi^T_1 i\tau_2 \Delta^{\dagger}  \Phi_1 + \mu_2 \Phi^T_2 i \tau_2 \Delta^\dagger \Phi_2 + \mu_3 \Phi^T_1 i\tau_2 \Delta^\dagger \Phi_2 + \mbox{H.c.} \right) \nonumber \\
&+ \left( \lambda_8 \Phi^\dagger_1 \Phi_1 + \lambda_9 \Phi^\dagger_2 \Phi_2 + (\lambda_{12} \Phi^\dagger_1 \Phi_2 + \mbox{H.c.})  \right) Tr \Delta^\dagger \Delta  \nonumber \\
& + \lambda'_8  \Phi^\dagger_1 \Delta \Delta^\dagger \Phi_1 + \lambda'_9 \Phi^\dagger_2 \Delta \Delta^\dagger \Phi_2 + (\lambda'_{12}  \Phi^\dagger_1 \Delta \Delta^\dagger \Phi_2 + \mbox{H.c.})\,. \label{eq:v2_c}
\end{align}
In terms of the Higgs basis, the minimal conditions for the  VEVs of $H_{1,2}$ and $\Delta$ can be obtained as:
\begin{subequations}
\begin{align}
& \left( \frac{c^4_\beta \lambda_1 + s^4_\beta \lambda_2}{2} + c^2_\beta s^2_\beta \lambda_{345} + 2 c^3_\beta s_\beta \lambda_6 + 2 c_\beta s^3_\beta \lambda_7 \right) v^2 = - c^2_\beta m^2_1 - s^2_\beta m^2_2 + 2 c_\beta s_\beta m^2_{12}  \nonumber \\
  & + \sqrt{2} v_\Delta \left( c^2_\beta \mu_1 + s^2_\beta \mu_2 + c_\beta s_\beta \mu_3 \right) - \frac{v^2_\Delta}{2} \left(c^2_\beta  \bar\lambda_8 + s^2_\beta \bar\lambda_9 \right)\,, \\
 & c_\beta s_\beta (m^2_2 -m^2_1) - m^2_{12} c_{2\beta} + \frac{v^2}{2} \left(-c^3_\beta s_\beta  \lambda_1 + c_\beta s^3_\beta \lambda_2 \right) + \frac{\lambda_{345} v^2}{2} c_\beta s_\beta c_{2\beta} + \frac{\lambda_6 v^2}{2} (-3c^2_\beta + c^4_\beta) \nonumber \\
 &+ \frac{\lambda_7 v^2}{2} (3c^2_\beta - s^4_\beta) = -\sqrt{2} c_\beta s_\beta v_\Delta (\mu_1 - \mu_2) + \frac{c_{2\beta} v_\Delta \mu_3 }{\sqrt{2}} + \frac{\bar\lambda_8 + \bar\lambda_9}{2} c_\beta s_\beta v^2_\Delta\,, \\
& \left( m^2_\Delta + \frac{\bar\lambda_8}{2} c^2_\beta v^2 + \frac{\bar\lambda_9}{2} s^2_\beta v^2  + (\lambda_{\Delta 1} + \lambda_{\Delta 2}) v^3_\Delta \right) v_\Delta = \frac{v^2}{2} \left( c^2_\beta \mu_1 + s^2_\beta \mu_2 + c_\beta s_\beta \mu_3 + s_{2\beta} \bar\lambda_{12}\right)\,,
 \end{align}
 \end{subequations}
where $\bar\lambda_{8}=\lambda_8 + \lambda'_8$, $\bar\lambda_{9}=\lambda_9 + \lambda'_9$, $\bar\lambda_{12} = \lambda_{12} + \lambda'_{12}$, and   the VEV of $\Delta$ is denoted by $v_\Delta$.  These relations are useful to simplify the expressions of scalar masses and trilinear scalar couplings. %
{ We note that since no discrete symmetry is imposed  in the 2HDM, both $H_{1,2}$ Higgs doublets are indistinguishable. It is simpler to directly use the Higgs basis in the scalar potential.   To compare our results with those given in Ref.~\cite{Ouazghour:2018mld}, here we employ the generic Higgs flavor basis, which is used in Ref.~\cite{Ouazghour:2018mld}.  Nevertheless, we show the more compact expressions  in appendix~\ref{app:SP_HB}.  }

If we drop the small effect from $(\lambda_{\Delta 1} + \lambda_{\Delta 2})v^3_\Delta$, the  Higgs triplet VEV can be obtained as:
\begin{equation}
v_\Delta \approx \frac{v^2}{\sqrt{2}} \frac{c^2_\beta \mu_1 + s^2_\beta \mu_2 + c_\beta s_\beta \mu_3}{\tilde{M}^2_\Delta}\,, \label{eq:v_D}
\end{equation}
with $\tilde{M}^2_\Delta = m^2_\Delta + (\bar\lambda_8  c^2_\beta v^2 + \bar\lambda_9 s^2_\beta v^2 + \bar\lambda_{12} s_{2\beta} )v^2 /2$. 
Since  $v_\Delta$ is bounded by the electroweak precision measurement,  and with the exception of the neutrino mass, its effect is irrelevant to the current study.  Precision measurement of the electroweak $\rho$ parameter gives a constraint that $v_\Delta \alt 8$~GeV~\cite{PDG}. To illustrate the importance of trilinear scalar couplings between the two Higgs doublets and the Higgs triplet on the muon $g-2$, we take $v_\Delta \sim O(10^{-3} -10^{-4})$~GeV.  { The considered parameter region can be easily achieved.  For instance, using $c^2_\beta \mu_1 + s^2_\beta \mu_2 + c_\beta s_\beta \mu_3 \approx  10^{-3}$ GeV and $\tilde{M}_\Delta = 500$~GeV, we obtain $v_\Delta \approx 1.7\times 10^{-4}$~GeV. }  As a result, the mass mixings of scalars, pseudoscalars, and charged scalars between $\Phi_i$ and $\Delta$ are phenomenologically negligible, justifying our earlier assumption.  Due to the doublet-triplet coupling terms  in Eq.~(\ref{eq:v2_c}), the $G^{0,\pm}$ and $H^\pm(A^0)$ in the Higgs basis of 2HDM are not the Goldstone modes and the physical states, and  the $(G^0, A^0, \eta^0)$ and $(G^\pm, H^\pm, \delta^\pm)$ states  will respectively mix. The only nonmixing  state is the doubly-charged Higgs, where from Eq.~(\ref{eq:v2_b}),  its mass can be expressed as:
   \begin{equation}
 m^2_{\delta^{\pm\pm}} = m^2_\Delta + \lambda_{\Delta 1} v^2_\Delta + \frac{v^2}{2} \left(c^2_\beta \lambda_8 + s^2_\beta \lambda_9 + \lambda_{12} s_{2\beta} \right)\,.
 \end{equation}
It can be seen that the new doublet-triplet couplings terms shift the $\delta^{\pm\pm}$ mass.

The detailed discussions for the scalar, pseudoscalar, and charged scalar mass matrix are given in appendix~\ref{app:mass_matrix}. We summarized the characteristics  as follows:  from Eqs.~(\ref{eq:CP-odd_M}) and  (\ref{eq:Charged_M}), it can be seen that $m^2_{G^0}$ and $m^2_{G^\pm}$ are proportional to $v^2_\Delta$. With $v_{\Delta}\sim O(10^{-4})$ GeV, their values can be dropped.  If  $\mu_3 = \tan2\beta (\mu_1 - \mu_2)$ is required,  we can find that the mixing matrix elements of $G^0A^0$ and  $G^- H^+$ become $O(v^2_\Delta)$, and the $G^0(A^0) \eta^0$, $G^-(H^-)\delta^+$, and $h(H)\delta^0$ matrix elements are of $O(v_\Delta)$. In comparison with the mass-square elements of other massive particles,  their mixing effects are small. Although the small mixing effects can have important influence on some processes, e.g. $\delta^{\pm\pm}\to W^\pm H^\pm$ can be induced, their influence on the muon $g-2$   can be indeed neglected. Hence, when we numerically estimate the muon $g-2$, we take $h(H)$, $H^\pm(\delta^\pm)$ and $A^0$ as the physical states; however, for other processes, one can take the mixing effects into account if necessary.

The new doublet-triplet couplings can cause the triplet scalar mass splittings,  and the mass differences  can be found as:
 \begin{align}
 m^2_{\delta^\pm} - m^2_{\delta^{\pm\pm}}  & = \frac{ v^2_\Delta}{2}  \lambda_{\Delta 2} + \frac{v^2}{4} \left( c^2_\beta \lambda'_8 + s^2_\beta \lambda'_9 + \lambda'_{12} s_{2\beta} \right)\,, \nonumber \\
m^2_{\delta^0}  - m^2_{\delta^{\pm\pm}}   &= (2 \lambda_{\Delta 1} + 3\lambda_{\Delta 2}) v^2_\Delta
 + \frac{v^2}{2} \left( c^2_\beta \lambda'_8 + s^2_\beta \lambda'_9  + \lambda'_{12} s_{2\beta}\right)\,. \label{eq:mass_dif_D}
 \end{align}
 It can be seen that the mass split can be or be less than  $O(100)$ GeV .

The trilinear interactions among a neutral Higgs boson and two charged Higgs bosons are given by:
 \begin{equation}
 {\cal L}_{H^0_i SS}=  -v\left[ \lambda_{H^0_i \delta^{--}\delta^{++} }  H^0_i \delta^{--} \delta^{++}  +  \lambda_{H^0_i \delta^{-}\delta^{+} }  H^0_i \delta^{-} \delta^{+}  +  \lambda_{H^0_i H^- H^+ }  H^{0}_i H^{-}  H^{+}  \right]\,,
 \end{equation}
  where the couplings are written as:
 \begin{align}
 \begin{split}
 \lambda_{H^0_1 \delta^{--} \delta^{++} }
 =& \lambda_8 c^2_\beta + \lambda_9 s^2_\beta + \lambda_{12} s_{2\beta} \,,  \\
 \lambda_{H^0_2 \delta^{--} \delta^{++}} 
 = & (- \lambda_8 + \lambda_9) c_\beta s_\beta  + \lambda_{12} c_{2\beta} \,,  \\
 \lambda_{H^0_1 \delta^{-} \delta^{+} } 
 =& \left(\lambda_8 + \frac{\lambda'_8}{2} \right) c^2_\beta  + \left(\lambda_9 + \frac{\lambda'_9}{2} \right) s^2_\beta  + \left( \lambda_{12} + \frac{\lambda'_{12} }{2}\right) s_{2\beta}  \,,  \\
 \lambda_{H^0_2 \delta^{-} \delta^{+} } 
 =&- \left( \lambda_8 - \lambda_9 + \frac{\lambda'_8 - \lambda'_9}{2}\right) + \left( \lambda_{12} + \frac{\lambda'_{12} }{2}\right) c_{2\beta}  \,,  \\
 \lambda_{H^0_1 H^- H^+} 
 =&  [\lambda_1 + \lambda_2 - 2(\lambda_4 + \lambda_5)] c^2_\beta s^2_\beta 
 + \lambda_3 (c^4_\beta + s^4_\beta)v - 2 (\lambda_6 - \lambda_7) c_\beta s_\beta c_{2\beta} ~,   \\
 \lambda_{H^0_2 H^- H^+} 
 =&  -\lambda_1 c_\beta s^3_\beta  + \lambda_2 c^3_\beta s_\beta  - \lambda_{345} c_\beta s_\beta c_{2\beta} 
 -\lambda_6  (3c^2_\beta s^2_\beta -s^4_\beta) + \lambda_7  \left( c^4_\beta -3 c^2_\beta s^2_\beta \right) \,. \label{eq:trilinear}
\end{split}
\end{align}
In the alignment limit of $c_{\beta-\alpha}=0$, the $h$ and $H$ trilinear  terms can be easily obtained by the replacement of $H^0_1 \to h$ and $H^0_2 \to -H$. Therefore, the corresponding trilinear couplings have the relations: 
 \begin{align}
 \lambda_{h\delta^{--(-)}\delta^{++(+)}}  & =  \lambda_{H^0_1\delta^{--(-)}\delta^{++(+)}}\,, ~~  \lambda_{hH^-H^+} =\lambda_{H^0_1 H^- H^+}\,, \nonumber \\
  \lambda_{H\delta^{--(-)}\delta^{++(+)}}  & =  -\lambda_{H^0_2\delta^{--(-)}\delta^{++(+)}}\,, ~~  \lambda_{HH^-H^+} =- \lambda_{H^0_2 H^- H^+}\,.
 \end{align}
We note that the pseudoscalar $A^0$ does not couple to the charged scalars in the CP-conserving case.

\subsection{Yukawa interactions
\label{sec:Yukawa interactions}}

The most general Yukawa couplings in the model are given by:
\begin{align}
-{\cal L}_Y &=  \bar Q_L Y^d_1 D_R \Phi_1 + \bar Q_L Y^{d}_2 D_R \Phi_2
+ \bar Q_L Y^u_1 U_R \tilde \Phi_1 + \bar Q_L Y^{u}_2 U_R \tilde \Phi_2
 \nonumber \\
&+  \bar L Y^\ell_1 \ell_R \Phi_1 + \bar L Y^{\ell}_2 \ell_R \Phi_2 +  \frac{1}{2} L^T C{\bf y}^{\nu} \, i\tau_2 \,\Delta  L  + {\rm H.c.}\,, 
\label{eq:Yu}
\end{align}
where the flavor indices are suppressed, ${\bf y^\nu}$ is a symmetric matrix, $Q_L$ $(L)$ denotes the quark (lepton) doublets, $q_R$ $(\ell_R)$ denotes the quark (lepton) singlets, $Y^f_{1,2}$ with $f = u, d, \ell$ are respectively the Yukawa matrices for the up-type quarks, down-type quarks, and charged leptons, $C$ is the charge conjugation operator, and $\tilde \Phi_i \equiv i\tau_2 \Phi^*_i$ with $\tau_2$ being the Pauli matrix.

Since $\Phi_1$ and $\Phi_2$ simultaneously couple to each type of fermions, flavor-changing neutral currents (FCNCs) naturally arise at tree level. The FCNC effects are usually suppressed by introducing, for example, a $Z_2$ discrete symmetry~\cite{Glashow:1976nt}.  In this case, the 2HDM can be categorized into Type-I~\cite{Haber:1978jt,Hall:1981bc}, Type-II~\cite{Donoghue:1978cj,Hall:1981bc}, Type-X, and Type-Y~\cite{Barger:1989fj,Savage:1991qh,Grossman:1994jb,Akeroyd:1994ga}.  See Ref.~\cite{Branco:2011iw} for a detailed review.  In addition to the above-mentioned schemes in 2HDM, the tree-level FCNCs can also be avoided by imposing a certain relation between $Y_1^f$ and $Y_2^f$, where $f = u$, $d$, and $\ell$.  The A2HDS assumes the relation $Y^f_{2} =\xi_f Y^f_{1}$, where $\xi_f$ is a proportionality constant~\cite{Pich:2009sp}.  Alternatively, one may also impose the condition $Y^f_{2} = N_I Y^f_{1} N^\dagger_I$~\cite{Ahn:2010zza,Chen:2011wp,Chen:2016xju}, where the possible $N_I$ matrices can be found in Ref.~\cite{Ahn:2010zza}.  In this work, we are considering the A2HDS.

With the assumed VEVs of $\Phi_i$, $Y^f_1$ and $Y^f_2$ in Eq.~(\ref{eq:Yu}) can be linearly combined to form two matrices:
\begin{align}
X_f &= c_\beta Y^f_1 + s_\beta Y^f_2 \,,  \nonumber \\
Z_f & = -s_\beta Y^f_1 + c_\beta Y^f_2\,,
\end{align}
so that $X_f(Z_f)$ is associated with the doublet $H_{1(2)}$, and  the fermion mass matrix can be obtained as  $M_f = X_f v/\sqrt{2}$.  Moreover, $M_f$ can be diagonalized by the unitary matrices $U^f_{L,R}$ in the way $M^{\rm dia}_{f} = U^f_L M_f U^{f \dag}_{R}$.  If $Y^f_1$ and $Y^f_2$ are two linearly independent matrices and cannot be diagonalized simultaneously, then tree-level FCNCs can arise due to the $Z_f$ couplings because its off-diagonal elements cannot be removed when $X_f$ is diagonalized.  When the A2HDS relation $Y^f_2 = \xi_f Y^f_1$ is taken, the Yukawa matrix can be related to the mass matrix as:
\begin{align}
 Y^f_1 & = \frac{\sqrt{2}}{c_\beta v} \frac{1}{1+ \xi_f t_\beta} M_f\,. 
\end{align}
As a result, both $Y^f_1$ and $Y^f_2$ now can be diagonalized simultaneously, and the  $H^0_{1,2}$ FCNCs are suppressed at the tree level.

For simplicity, we only concentrate on the CP-conserving case and assume $\xi_f$ to be real, though they can generally be complex.  Using Eqs.~(\ref{eq:Higgs_Basis}) and $(\ref{eq:2HDM_mass_Basis})$ and the notations used in~\cite{Pich:2009sp}, the mass terms and Yukawa interactions with $h$, $H$, $A^0$, and $H^\pm$ are found to be:
 \begin{align}
-{\cal L}_Y 
=& \sum_{f=u,d,\ell} \left[ \bar f_L M^{\rm dia}_f f_R  +  \frac{c_{\beta-\alpha} - \zeta_f s_{\beta-\alpha}}{v} \bar f_L M^{\rm dia}_{f} f_R  H + \frac{s_{\beta-\alpha} + \zeta_f c_{\beta-\alpha}}{v} \bar f_L M^{\rm dia}_{f} f_R  h \right] 
\nonumber \\
& + \sum_{f=u,d,\ell}  \frac{\zeta_f s_f }{v}\bar f_L M^{\rm dia}_f f_R (i A^0)  + \frac{\sqrt{2}}{v} \bar u \left( V_{\rm CKM} \zeta_d M^{\rm dia}_{d} P_R -  M^{\rm dia}_{u} \zeta^*_u V_{\rm CKM} P_L \right) d H^+  
\nonumber \\
& + \frac{\sqrt{2}}{v} \bar\nu \left( V_{\rm PMNS} \zeta_\ell M^{\rm dia}_{\ell} P_R \right) \ell H^+ + \mbox{H.c.}
\,, \label{eq:L_Y}
\end{align}
where $s_{d,\ell}=+1$, $s_{u}=-1$; $P_{R(L)}=(1\pm \gamma_5)/2$ are the chirality projection operators, $V_{\rm CKM}= V^u_L V^{d\dag}_L$ is the Cabibbo-Kobayashi-Maskawa (CKM) matrix~\cite{Cabibbo:1963yz, Kobayashi:1973fv}, $V_{\rm PMNS} = V^\nu_L V^{\ell\dag}_L$ is the Pontecorvo-Maki-Nakagawa-Sakata (PMNS) matrix ~\cite{Pontecorvo:1957cp,Maki:1962mu}, and
 \begin{equation}
\zeta_f = \frac{\xi_f - t_\beta}{1+ \xi_f t_\beta}\,,
 \end{equation}
with $t_\beta = \tan\beta$.  In general, $\zeta_f$ can be complex numbers as $\xi_f$~\cite{Pich:2009sp},  and their magnitudes can be large without requiring a large $\tan\beta$.  In this study, we only focus on the CP-conserving case.  The $A^0$ and $H^\pm$ Yukawa couplings do not depend on $c_{\beta-\alpha}$ $(s_{\beta-\alpha})$. 
For comparison, we show in Table~\ref{tab:types} the vanishing and non-vanishing $Y^f_{1,2}$ for various 2HDM types and the associated $\zeta_f$.  In particular, $\zeta_f$ in Type-I, -II, -X, and -Y can be obtained from the A2HDS by taking an appropriate limit of $\xi_f$ and thus $\zeta_f$:
\begin{align}
  Y^f_2 &=0: ~ \xi_f =0 \,~~\rightarrow~ \zeta_f = - t_\beta\,, \nonumber \\
  Y^f_1 &=0: ~ \xi_f = \infty ~ \rightarrow~ \zeta _f =t^{-1}_\beta\,. 
\end{align}

\begin{table}[thp]
\caption{ Vanishing (mark by 0) and non-vanishing (marked by $\times$) Yukawa matrices of various 2HDM types and the corresponding $\zeta_f$. }
\begin{center}
\begin{tabular}{c|ccccccccc} \hline \hline
  & ~~$Y^{d}_1$~~ & ~~$Y^{u}_{1}$~~  &  ~~$Y^{\ell}_{1}$~~ & ~~$Y^{d}_2$~~ & ~~$Y^{u}_{2}$~~  &  ~~$Y^{\ell}_{2}$~~  & ~~$\zeta_u$~~ & ~~$\zeta_d$~~  &  ~~$\zeta_\ell$~~  \\ \hline
  Type I  & 0 & 0 & 0 & $\times$ & $\times$ & $\times$ & $t^{-1}_\beta$ & $t^{-1}_\beta$ & $t^{-1}_{\beta}$ \\ \hline
  Type II  &  $\times$ & $0$ & $\times$ & $0$ & $\times$ & $0$ & $t^{-1}_\beta$ & $-t_\beta$ & $-t_\beta$ \\ \hline
  Type X & $0$ & $0$ & $\times$ & $\times$ & $\times$ & $0$ & $t^{-1}_\beta$ & $t^{-1}_\beta$ & $-t_\beta$\\ \hline
  Type Y &  $\times$ & $0$ & $0$ & $0$ & $\times$ & $\times$ & $t^{-1}_\beta$ & $-t_\beta$ & $t^{-1}_\beta$\\ \hline
  A2HDS & $\times$ & $\times$ & $\times$ & $\times$ & $\times$ & $\times$ & $\displaystyle \frac{\xi_u - t_\beta}{1+ \xi_u t_\beta}$ 
  & $\displaystyle \frac{\xi_d - t_\beta}{1+ \xi_d t_\beta}$ & $\displaystyle \frac{\xi_\ell - t_\beta}{1+ \xi_\ell t_\beta}$ \\ \hline \hline  
\end{tabular}
\end{center}
\label{tab:types}
\end{table}%

Since the SM-like Higgs couplings generally depend on $c_{\beta-\alpha}$ $(s_{\beta-\alpha})$ and $\zeta_f$, the current Higgs production and decay measurements put stringent constraints on the value of $c_{\beta -\alpha}$.  Here we simply take the alignment limit with $\beta-\alpha\approx \pi/2$~\cite{Gunion:2002zf}, i.e., $c_{\beta-\alpha} \to 0$ ($s_{\beta -\alpha} \to 1 $).  As a result, the $H$ and $A^0$ couplings to the SM fermions have the same magnitude and are dictated by $\zeta_f$.  In this work, we demonstrate how a large $\zeta_\ell$ can affect the muon $g-2$ when $m_H > m_h$ and the $H\delta^{++} \delta^{--}$ and $H\delta^{+} \delta^{-}$ couplings are present.

Using the component fields of the Higgs triplet shown in Eq.~(\ref{eq:scalars}), the neutrino mass and lepton Yukawa interactions with the triplet fields are given by:
\begin{align}
\label{eq:yu_neutrino}
{\cal L}_Y \supset  \frac{1}{2}\overline {\nu^C_L} M_\nu \nu_L +  \frac{1}{2}\overline {\nu^C_L} M_\nu \nu_L  \frac{\delta^0 + i\eta^0}{v_\Delta} - \overline{\nu^C_L} \frac{{\bf y^\nu} }{\sqrt{2}} \ell_L \delta^{+} - \frac{1}{2} \overline{\ell^C_L} {\bf y^\nu} \ell_L \delta^{++} 
+ \mbox{H.c.}\,,
\end{align}
where $f^C=C\gamma^0 f^*$ and  $M_\nu= {\bf y^\nu} v_\Delta/\sqrt{2}$ is the neutrino mass matrix.  In order to fit the neutrino data, the values of $(M_\nu)_{ij}$ has to be of $O(10^{-3}-10^{-2})$ eV~\cite{Chen:2014xva,Chen:2017gvf,Chen:2020ptg}.  In the type-II seesaw model with the assumed triplet VEV $v_\Delta \sim {\cal O}(10^{-3} - 10^{-4})$~GeV, the Yukawa couplings ${\bf y}^\nu_{ij}$ are very small, $\alt {\cal O}(10^{-7})$.  Therefore, $\delta^\pm$ and $\delta^{\pm\pm}$ of ${\cal O}(10^2)$~GeV mass have negligible effects on most lepton processes.


As we will numerically show below, the Yukawa couplings as well as the trilinear scalar couplings $ H \delta^{--(-)} \delta^{++(+)}$ and $HH^{-}H^{+}$, arising from the scalar potential, play important roles in producing a sizable correction to the muon $g-2$.

\section{One- and two-loop muon $g-2$} \label{sec:loop_g2}

\begin{figure}[phtb]
\begin{center}
\includegraphics[scale=0.75]{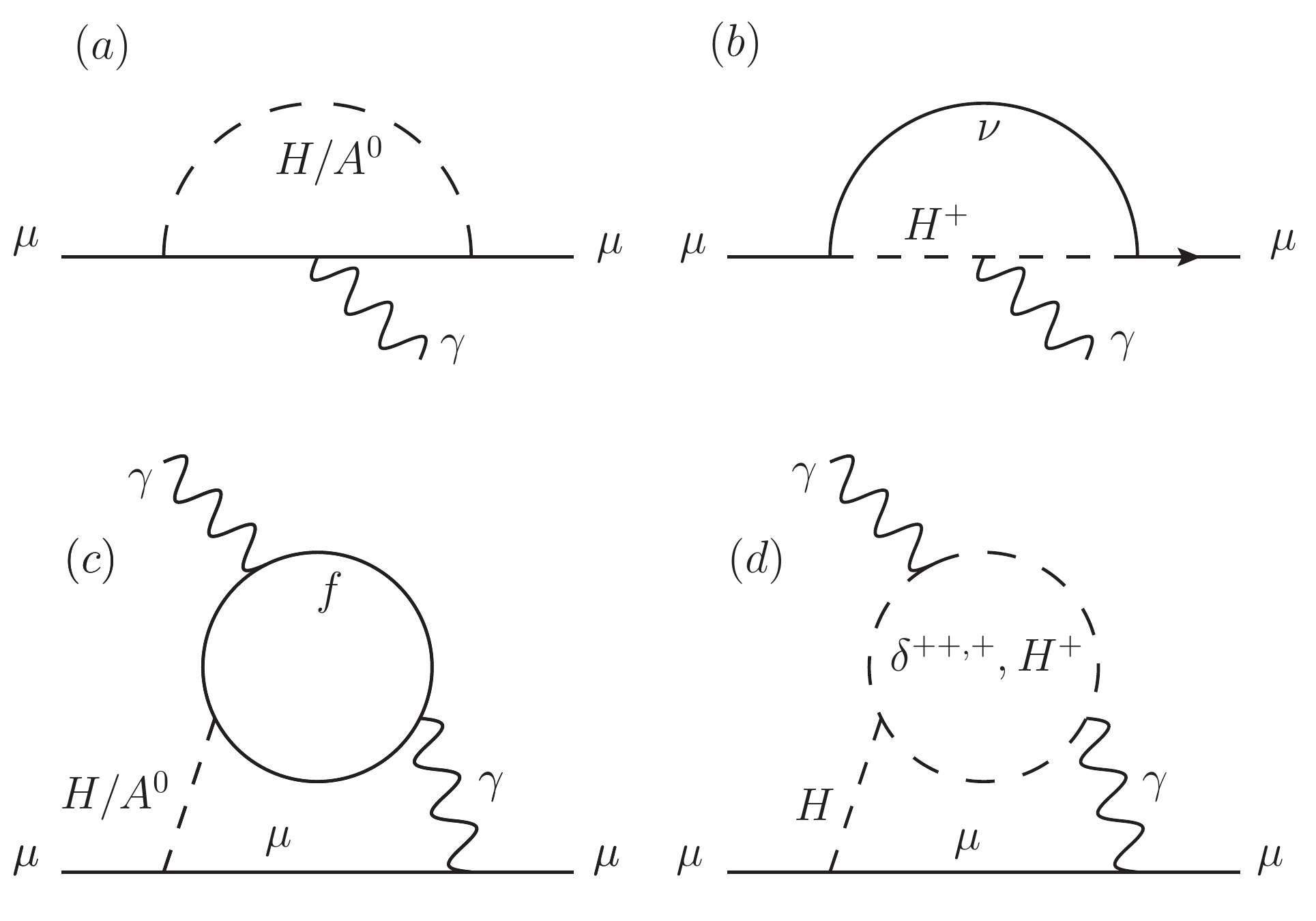}
\caption{One-loop and two-loop Barr-Zee type Feynman diagrams for the muon $g-2$, where $f$ in plot (c) includes top (bottom) quark and $\tau$ lepton. }
\label{fig:1_2loop_Feyn}
\end{center}
\end{figure}

The electromagnetic interaction of a lepton can be written as:
\begin{equation}
 \overline \ell (p') \Gamma^\mu \ell(p)
 = \overline \ell (p') \left[ \gamma^\mu  F_1(k^2) + \frac{i \sigma^{\mu \nu} k_\nu}{2m_\ell} F_2(k^2) \right] \ell(p) \,. \label{eq:mg2}
\end{equation}
The lepton anomalous magnetic dipole moment is then defined by
\begin{equation}
a_\ell = \frac{g_\ell-2}{2} = F_2(0)\,.
\end{equation}
Since the magnetic moment is associated with dipole operator, the lepton $g-2$ originates from radiative quantum corrections.  In the model, the one-loop corrections from new physics are induced by the mediation of $H$, $A^0$, and $H^\pm$, where the associated Feynman diagrams are shown in Fig.~\ref{fig:1_2loop_Feyn}(a) and (b). 
Moreover, it is known that the two-loop Barr-Zee type diagrams can have important contributions to the magnetic dipole moment due to a large coupling enhancement~\cite{Barr:1990vd,Chang:2000ii}.  The potentially large two-loop diagrams mediated by heavy fermions, including top, bottom, and $\tau$, are shown in Fig.~\ref{fig:1_2loop_Feyn}(c).  The essential mechanism contributing to the muon $g-2$ in the model is the two-loop with Barr-Zee type diagram mediated by the charged scalars, including $\delta^{++}$, $\delta^{+}$, and $H^+$, as shown in Fig.~\ref{fig:1_2loop_Feyn}(d).  In addition to the lepton Yukawa coupling, such diagrams further enjoy the enhancement of the electric charges associated with the charged scalars.

The one-loop corrections to the anomalous magnetic dipole moment in the 2HDM have been studied long time ago~\cite{Leveille:1977rc,Haber:1978jt,Krawczyk:1996sm,Dedes:2001nx}.  Using the Yukawa couplings shown in Eq.~(\ref{eq:L_Y}), the muon $g-2$ from Fig.~\ref{fig:1_2loop_Feyn}(a) and (b) can be expressed as:
\begin{align}
\label{eq:1loop_g2}
\Delta a^{1,H/A^0}_\mu &= \frac{ m^2_\mu  }{8\pi^2 v^2} \left[ (y^H_\ell)^2 \int^1_0 \frac{r^\mu_H x^2 (2-x) }{1-x(1- r^\mu_H x)}
- (y^{A^0}_\ell)^2\int^1_0 \frac{r^\mu_{A^0}  x^3 }{1-x(1- r^\mu_{A^0}  x)} \right]\,,\nonumber \\
 \Delta a^{1,H^\pm}_\mu &= \frac{ m^2_\mu \zeta^2_\ell }{8\pi^2 v^2}  \int^1_0 \frac{r^\mu_{H^\pm} x^2 (1-x) }{1-x(1- r^\mu_{H^\pm} x)}  \,,
\end{align}
where $r^{f}_{B} =m^2_f/m^2_{B}$ with $B = H,~ A^0,~ H^\pm$ and the Yukawa couplings $y^{H,A^0}_f$ are defined as:
\begin{equation}
y^H_f = s_{\beta-\alpha} \zeta_f -c_{\beta-\alpha}\,,~~ y^{A^0}_f=- s_f \zeta_f\,.
\end{equation}
From the expressions, it can be seen that the induced muon $g-2$ is proportional to $m^4_\mu$.  One of the four factors of $m_\mu$ comes from the definition in Eq.~(\ref{eq:mg2}), another comes from the mass insertion for chirality flip, and the rest two enter through the two Yukawa interaction vertices, each of which is proportional to the muon mass.  Thus, if the intermediate scalar mass is of ${\cal O}(100)$~GeV, the resulting muon $g-2$ is far below $10^{-9}$.  To get $\Delta a_\mu$ up to $\sim 10^{-9}$, the mediating particle has to be as light as tens of GeV.  This was the observation previously found for the 2HDM in the literature.

Following the results shown in Ref.~\cite{Ilisie:2015tra}, the two-loop Barr-Zee type diagrams with fermion and charged scalars can be written as:
\begin{align}
\label{eq:2loop_g2}
\Delta a^{2,f}_\mu &= \sum_{f=t,b,\tau} \frac{ \alpha_{\rm em} m^2_\mu  N^f_C Q^2_f }{4\pi^3 v^2} \left[ y^{H}_f y^{H}_\ell J^{H}_{f} \left(r^f_H \right) +  y^{A^0}_f  y^{A^0}_\ell J^{A^0}_{f} \left( r^f_{A^0} \right) \right]\,,\nonumber \\
\Delta a^{2,S}_\mu &=\sum_{S=\delta^{++},\delta^{+}, H^{+}} \frac{\alpha_{\rm em}  Q^2_S r^\mu_{H}}{8\pi^3 }  \zeta_\ell \lambda_{HS^* S} \, J_{S}(r^{S}_{H})   \,,
\end{align}
where $N^f_C$ is the number of color for the fermion $f$, $Q_P$ ($P = f,~ S$) is the electric charge of the particle, and the loop functions are given by:
  \begin{align}
  \label{eq:loop_fun}
  J^{H}_{f} (z) &= \frac{z}{2} \int^1_0 \frac{2 x (1-x) -1}{z-x(1-x)}  \ln\left( \frac{z}{x(1-x)}\right)\,, \nonumber \\
   J^{A^0}_{f} (z) &= \frac{z}{2} \int^1_0 \frac{1} {z-x(1-x)}  \ln\left( \frac{z}{x(1-x)}\right)\,, \nonumber \\
   J_S(z) & =  \frac{1}{2} \int^1_0 \frac{ x(1-x)}{z-x(1-x)}  \ln\left( \frac{z}{x(1-x)}\right)\,.
  \end{align}
The two-loop results are proportional to $m^2_\mu$ because there is only one muon Yukawa coupling involved.  It can be seen that when $y^{H/A}_{u,d}$ are strictly bounded by the experimental data, their contributions become subleading, and $\Delta a^{2,S}_\mu$ is the dominant effect.

\section{Numerical Analysis}\label{sec:NA}

In this section, we present how we choose the parameters in our model, how they affect $\Delta a_\mu$ at one-loop and partial two-loop levels, and how the doubly-charged Higgs phenomenology at the LHC is modified.

\subsection{Parameter choice} 

Among the parameters in the Yukawa  and scalar sectors, most relevant ones for the muon $g-2$ are combinations of the Yukawa matrix elements, the quartic scalar couplings, and $t_\beta$ that appear in various couplings.  More explicitly, the relevant parameters are: $c_{\beta-\alpha}~(s_{\beta-\alpha})$, $\zeta_{u,d,\ell}$, $m_{H,A^0,H^\pm}$, $m_{\delta^{\pm\pm}, \delta^{\pm}}$, $\lambda_{H\delta^{--}\delta^{--}}$,  $\lambda_{H\delta^{-}\delta^{-}}$, and $\lambda_{HH^{-}H^{+}}$.  We will show how they contribute the muon $g-2$.

Before a numerical analysis, we first need to find the allowed parameter space for the model.  All potential constraints from experimental measurements, including various flavor physics processes, Higgs data, and electroweak precision observables, and theoretical bounds, such as perturbative unitarity and positivity of the scalar potential, have to be taken into account.   Recently, such a global fit, considering the theoretical constraints has been done in the A2HDS~\cite{Eberhardt:2020dat}.  In this work, we will follow the global fit results in Ref.~\cite{Eberhardt:2020dat} when the parameter values are taken for the numerical estimations.

Two scenarios, the light scenario and the heavy scenario, are discussed by in Ref.~\cite{Eberhardt:2020dat}, where the former refers to the case with $m_H > m_h$ and the latter has $m_H < m_h$.  Since we are interested in the heavy scalar boson contribution to $\Delta a_\mu$, we will concentrate on the light scenario.

The values of parameters used in our numerical analysis are described below.  Using the experimental data at $2\sigma$ errors, the global fit gives $|c_{\beta-\alpha}|< 0.04$.  Thus, we will take the alignment limit of $c_{\beta-\alpha}=0$.  Under this limit, the $HW^-W^+$ and $HZZ$ couplings vanish identically.  It is found that $\zeta_{u,d}$ have to be of the same sign and their values are restricted to small-value regions when $|\zeta_\ell|$ approaches the boundary of maximum, i.e., $|\zeta_\ell |=100$.  Moreover, the sign of $\zeta_\ell$ cannot be determined by the global fit, and it always appears in the product along with other parameters  in $\Delta a_\mu$, e.g., $\zeta_\ell \lambda_{H\delta^{--} \delta^{++}}$.  We can thus fix the value of $\zeta_\ell$ and let the associated parameter vary.  In numerical calculations, we take:
 \begin{align}
 \zeta_{u}= 0.1,~\zeta_{d}=10,~\zeta_{\ell}=-100\,.
 \end{align}
Since the maximally allowed value of $|\zeta_\ell|$ in the negative region is larger than that in positive region, we assume $\zeta_\ell$ to be negative.  The signs of $\zeta_{u,d}$ are taken to fit the positive $\Delta a_\mu$.

From Eq.~(\ref{eq:trilinear}), it is seen that the trilinear couplings have involved relations with the parameters in the scalar potential.  If we assume that the 2HDM with a $Z_2$ symmetry contributes little to the $HH^-H^+$ coupling, the $\lambda_{6,7}$ parameters in A2HDS become the dominant source, i.e.,
 \begin{equation}
 \lambda_{HH^-H^+} \simeq \lambda_6 \left(3 c^2_\beta s^2_\beta -s^4_\beta \right) - \lambda_7 \left( c^4_\beta - 3 c^2_\beta s^2_\beta \right) \,.
 \end{equation}
Since the constrained $\lambda_{6,7}$ values allow $\lambda_6\simeq -3.5$ and $\lambda_7 \simeq -2.5$, we will take $\lambda_{HH^-H^+} \approx 1.5$ to estimate the muon $g-2$. 

According to the results shown in Ref.~\cite{Ouazghour:2018mld}, the allowed values for $|\lambda_{8,9}|$ and $|\lambda'_{8,9}|$ can be of ${\cal O}(10)$\footnote{The parameters $\lambda_{8,9}$ and $\lambda'_{8,9}$ used in this paper correspond respectively to the parameters $\lambda_{6,7,8,9}$ in Ref.~\cite{Ouazghour:2018mld}. }, where the Higgs data and the theoretical constraints have been imposed.  With $c_\beta=s_\beta=1/\sqrt{2}$, it is expected that   $\lambda_{H\delta^{--(-)}\delta^{++(+)}} \lesssim 5$ can be conservative upper bounds.  The upper bound is consistent with that used in Ref.~\cite{Ilisie:2014hea} for the $HH^-H^+$ coupling due to the perturbative requirement.  For simplicity, we take $\lambda_{H\delta^{-}\delta^{+}}=\lambda_{H\delta^{--}\delta^{++}}$. Since $\zeta_\ell$ and $\lambda_{H\delta^{--}\delta^{++}}$ show up together in $\Delta a_\mu$, we can consider $\zeta_\ell \lambda_{H\delta^{--} \delta^{++}}$ as a single variable because $\zeta_\ell=-100$ is fixed.

Global fits cannot determine the masses of the involved new scalar particles.  Nevertheless, their mass differences are strongly correlated and constrained.  In our numerical analysis, we take $m_{\delta^\pm}=m_{\delta^{\pm\pm}} + 100$~GeV according to Eq.~(\ref{eq:mass_dif_D}).  When discussing the CP-even or CP-odd scalar effects, we take $m_{H(A^0)}$ as a free parameter.  When combining the effects of $H$ and $A^0$ together, we take $m_H=m_{H^\pm}$ and $m_{A^0}=m_H \pm 50$~GeV.
 
{
The parameters $|\zeta_\ell|$ and  $m_{H^\pm}$ can also be bounded by the Michel parameters~\cite{Michel:1949qe} in the $\tau$ decays~\cite{Abe:2015oca,Crivellin:2015hha,Chun:2015hsa,Enomoto:2015wbn,Tobe:2016qhz}. The Michel parameters in the leptonic $\tau$ decay are defined as:
 \begin{align}
 \frac{d\Gamma_{\tau \to \ell \bar\nu_\ell \nu_\tau}}{dx}  & \propto 
 f_0(x) + \rho_\ell  f_1(x) +  \eta_\ell \frac{m_\ell}{m_\tau} f_2(x) 
- P_\tau \xi_\ell  \left( g_1(x) + \delta_\ell g_2(x)\right)\,,
 \end{align}
  where $x=E_\ell/E_{\ell_{\rm max}}$ and $E_{\ell_{\rm max}} = m_\tau (1 + m_\ell^2/m_\tau^2) / 2$, $P_\tau$ is the $\tau$-lepton polarization, and the explicit expressions of  functions $f_{i}$ can be found in~\cite{PDG}. In the SM, the Michel parameters are predicted as $\rho_\ell =3/4$, $\eta_\ell=0$, $\xi_\ell=1$, and $\delta_\ell=3/4$, whereas the current experimental values are: $\rho^{\rm exp}_\ell=0.749 \pm 0.008$, $\eta^{\rm exp}_\ell=0.015\pm 0.021$, $\xi^{\rm exp}_\ell=0.981\pm 0.031$, and $(\delta_\ell \xi_\ell)^{\rm exp} = 0.79 \pm 0.04$. We will concentrate on the $\eta_\ell$ and $\xi_\ell$ parameters because they are sensitive to the scalar couplings. 
  
  The general transition matrix element for $\tau\to \ell \bar\nu_\ell \nu_\tau$ can be written as~\cite{Fetscher:1986uj}:
  \begin{equation}
  {\cal M} = 4 \frac{G_F}{\sqrt{2}} \sum_{\substack{\kappa=S,V,T \\ \epsilon, \lambda=R,T}}  
   g^\kappa_{\epsilon \lambda} \langle  \bar\ell_\epsilon | \Gamma^\kappa | (\nu_\ell)_n \rangle \langle  (\bar \nu_\tau)_m | \Gamma_\kappa | \tau_\lambda \rangle \,, \label{eq:gen_int}
  \end{equation}
  where $\kappa=S,V,T$ denotes the type of interaction, $\epsilon (\lambda)=R,L$ is the lepton chirality, and  the chirality of $m(n)$ can be determined when   $\kappa$ and $\epsilon(\lambda)$  are fixed. In the SM, due to the $V-A$ interaction,  we only have $g^V_{LL}=-1$. Since the involved couplings in the A2HDM are scalar and vector types and the $H^\pm$-Yukawa coupling is proportional to the lepton mass, we only need to consider the muon mode and the effective couplings $g^S_{RR}$ and $g^V_{LL}$.  Thus, the Michel parameters of $\eta_\mu$ and $\xi_\mu$ are  expressed as:
   \begin{align}
   \eta_\mu & = \frac{1}{2} {\rm Re}( g^S_{RR} \, g^{V*}_{LL}) \,, \nonumber \\
   \xi_\mu & = |g^V_{LL}|^2 - \frac{1}{4} |g^S_{RR} |^2\,,
   \end{align}
   where $g^S_{RR}$ and $g^V_{LL}$ with one-loop corrections~\cite{Crivellin:2015hha} in the model are given by:
   \begin{align}
    g^S_{RR} & = \frac{m_\mu}{m_\tau} \left( \frac{m_\tau \zeta_\ell}{m_{H^\pm}}\right)^2 \,, \nonumber \\
   g^V_{LL} & = -1 - \frac{\zeta_\ell^2}{32 \pi^2} \frac{m^2_\tau}{v^2} \left( F\left(\frac{m^2_{A^0}}{m^2_{H^\pm}}\right) + F\left(\frac{m^2_{H}}{m^2_{H^\pm}} \right) \right)\,,
   \end{align}
with the loop function $F$ defined by
    \begin{equation}
    F(a) = \frac{1}{2} + \frac{1+a}{4 (1-a) } \ln a\,.
    \end{equation}
   To illustrate the constraints from the measured Michel parameters, we show the contours of $\eta_\mu$ (left panel) and $\xi_\mu$ (right panel) in the $|\zeta_\ell|$-$m_{H^\pm}$ plane in Fig.~\ref{fig:Michel}, where $m_{A^0}=50$~GeV and $m_{H}=100$~GeV are used.  The dashed line in the left plot corresponds to the $2\sigma$ lower bound of $\eta^{\rm exp}_\mu$, pointing to the lower right region as more favorable parameter space.  For example, $m_{H^\pm} \lesssim 180$~GeV is excluded when $|\zeta_\ell|=100$.  The right plot, on the other hand, does not show much constraining powers as the entire parameter space gives values consistent with $\xi^{\rm exp}_\mu$ at the $2\sigma$ level.  Therefore, to avoid the constraint from $\eta^{\rm exp}_\mu$, we take $m_{H^\pm} \gtrsim 180$~GeV in the following numerical analysis.

\begin{figure}[phtb]
\begin{center}
\includegraphics[scale=0.6]{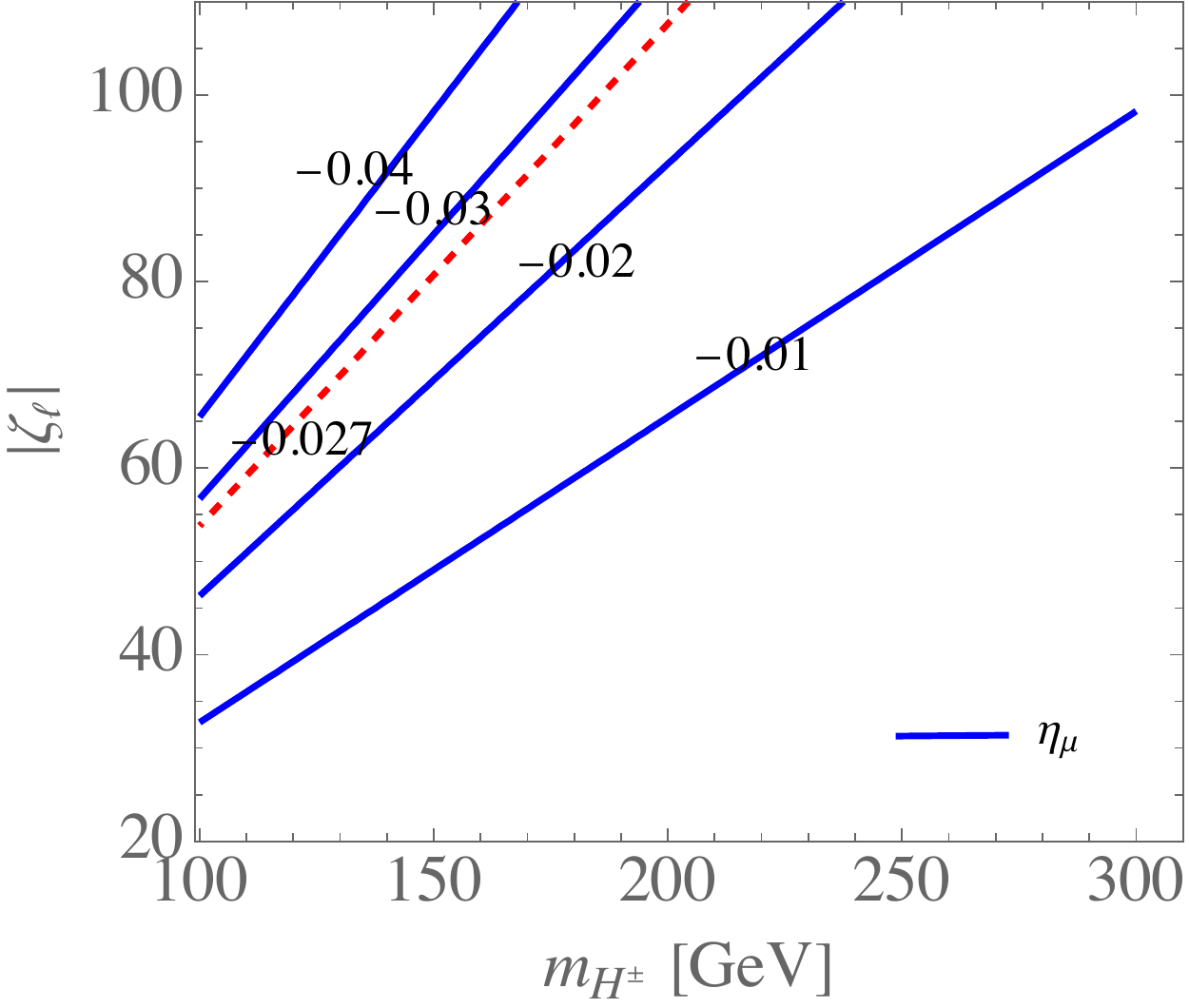}
\hspace{5mm}
\includegraphics[scale=0.6]{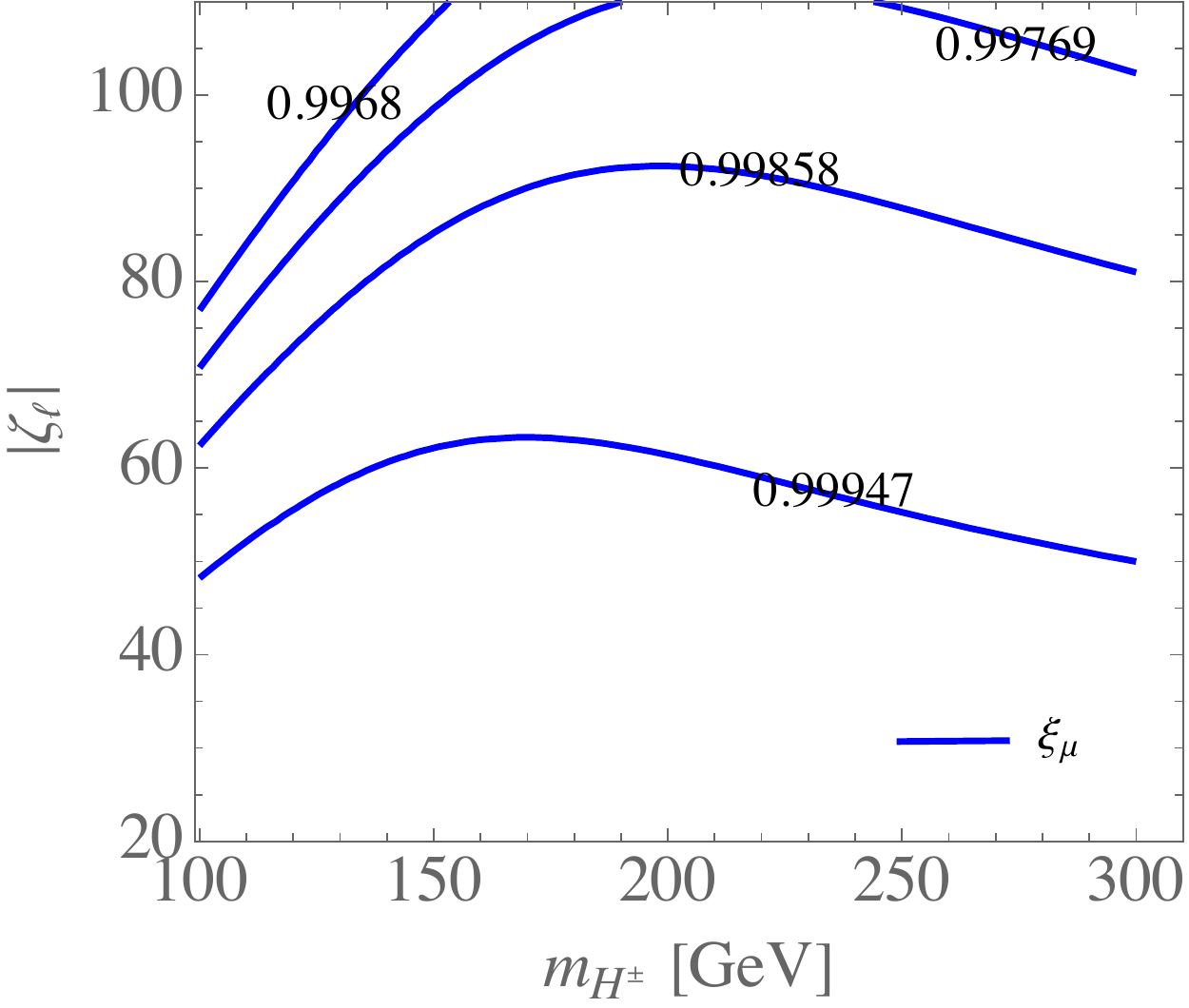}
 \caption{  Contours for the Michel parameters $\eta_\mu$ (left panel) and $\xi_\mu$ (right panel) as a function of $|\zeta_\ell|$ and $m_{H^\pm}$, where $m_{A^0}=50$~GeV and $m_{H}=100$~GeV are used. } \label{fig:Michel}
\end{center}
\end{figure}

}

\subsection{ Muon $g-2$}

In the following, we divide our discussion of muon $g-2$ into three contributing parts.

\subsubsection{One-loop contribution}

\begin{figure}[phtb]
\begin{center}
\includegraphics[scale=0.57]{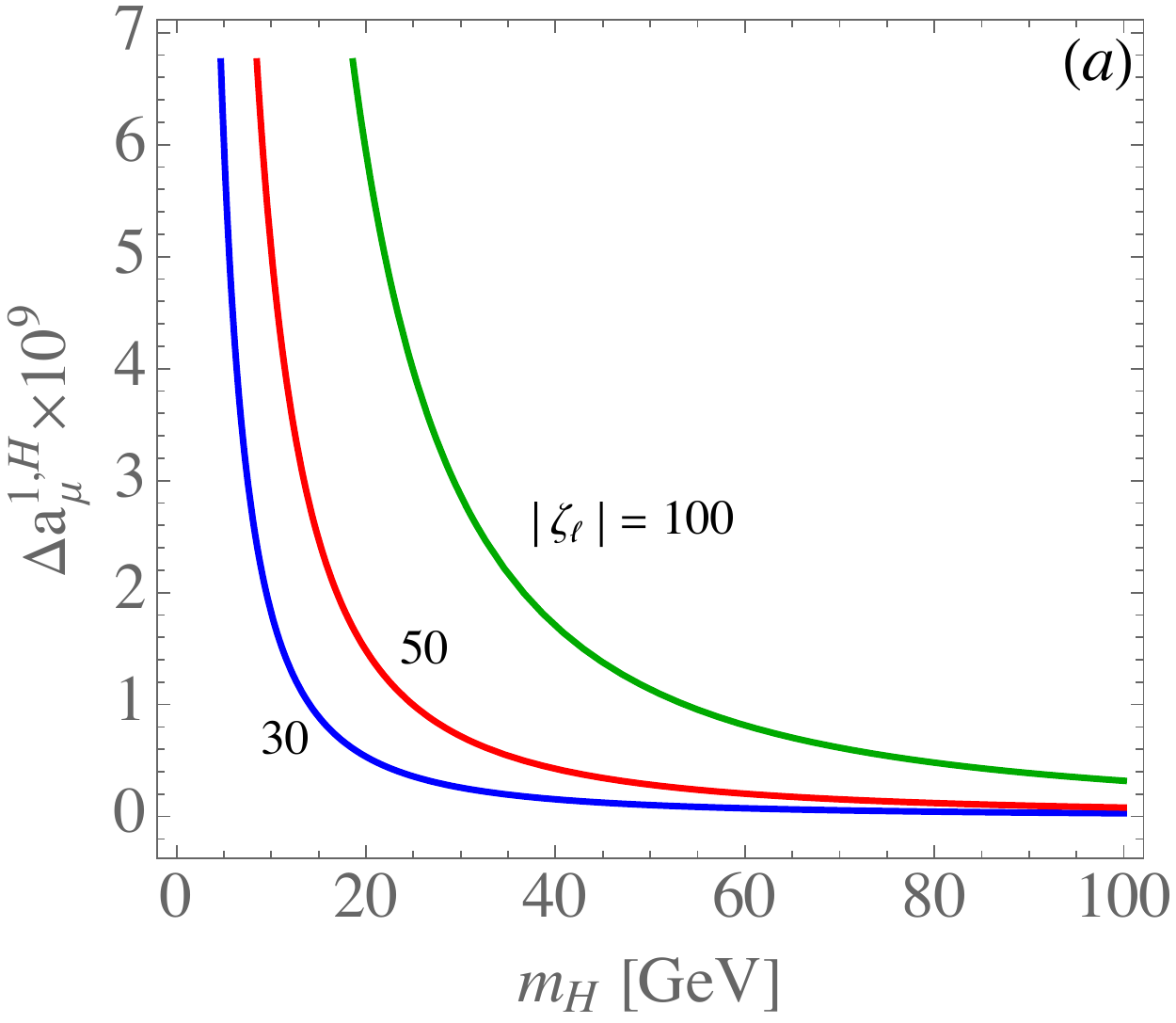}
\hspace{5mm}
\includegraphics[scale=0.6]{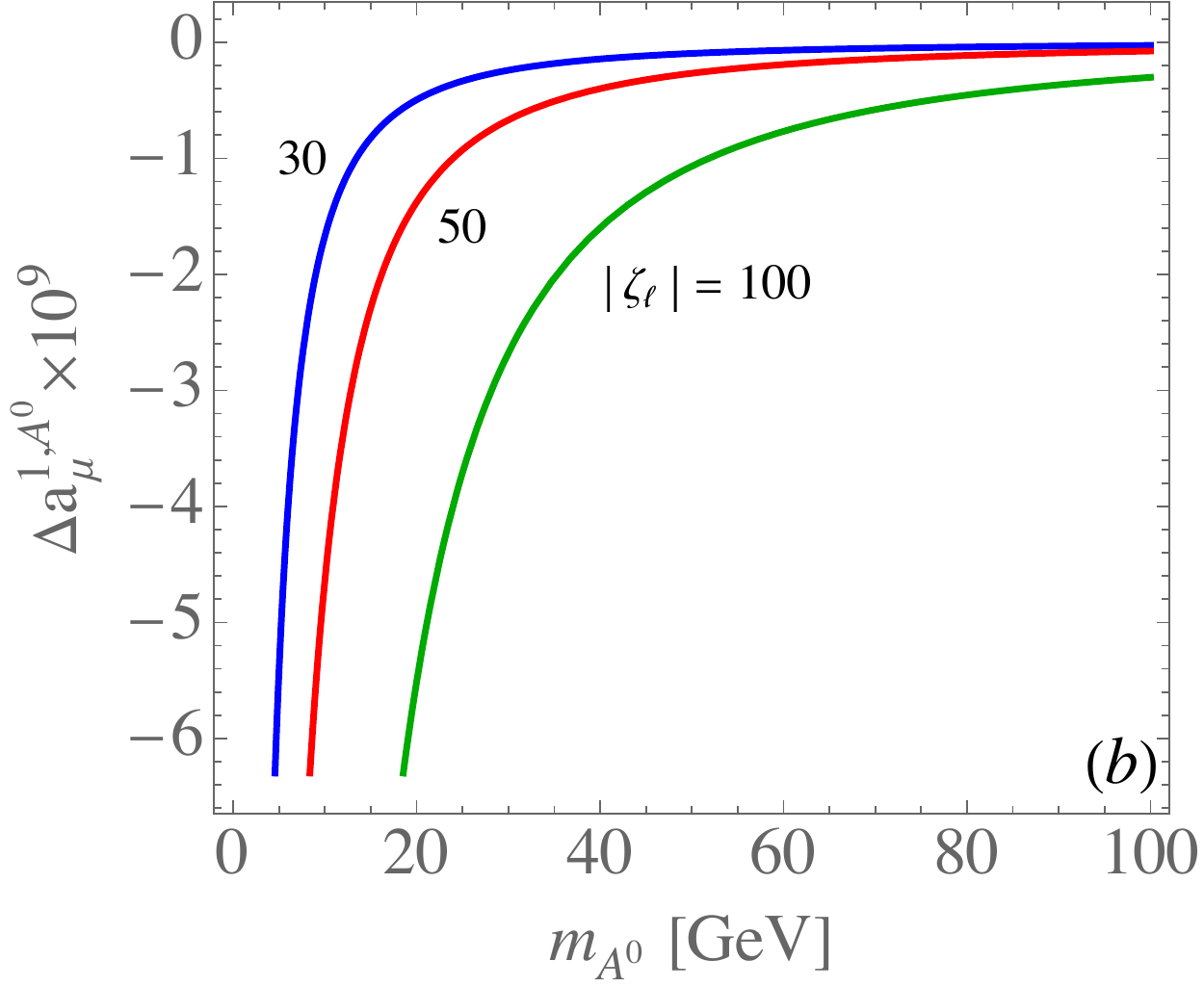}
\\
\includegraphics[scale=0.63]{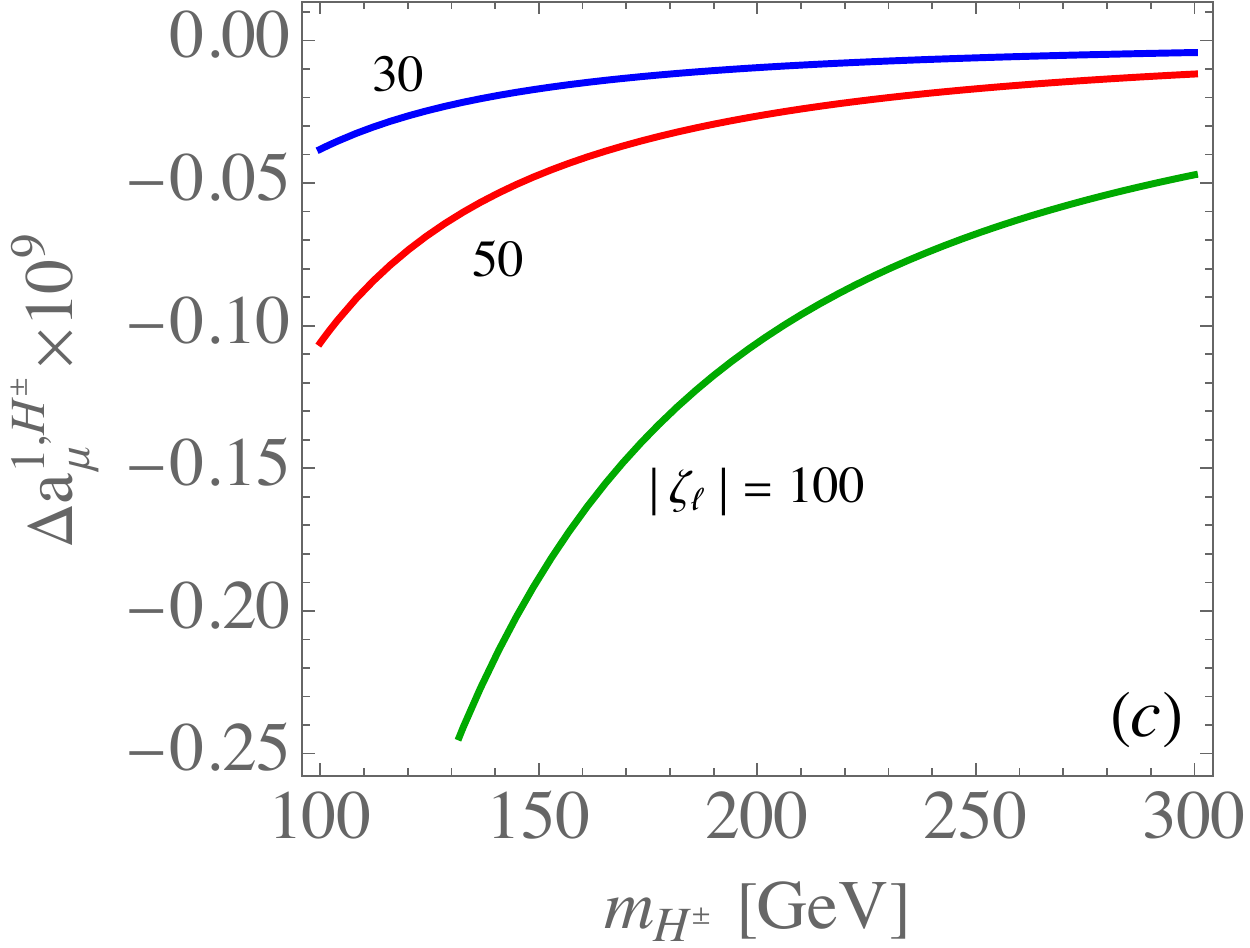}
\hspace{5mm}
\includegraphics[scale=0.57]{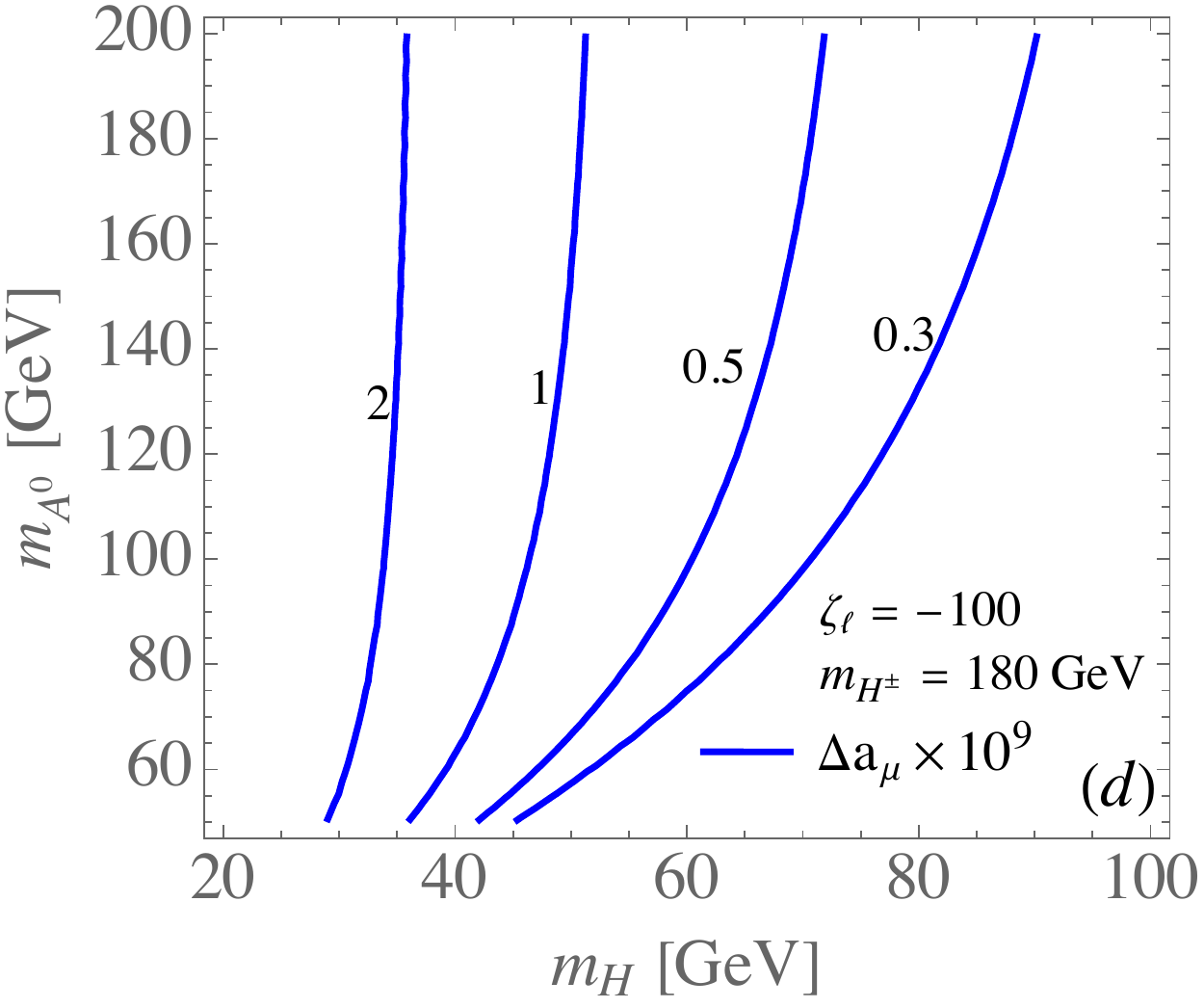}
 \caption{ One-loop muon $g-2$ within the 2HDM, induced by the mediation of (a) $H$, (b) $A^0$, and  (c) $H^\pm$, as a function of respectively $m_{H, A^0, H^\pm}$, where the curves correspond to $|\zeta_\ell|=(30, 50, 100)$.  (d) Contours of $\Delta a_\mu$ with the combination of $H^0$, $A^0$, and $H^\pm$, where  $\zeta_\ell=-100$ and $m_{H^\pm}=180$~GeV are used.  } \label{fig:1loop_g2}
\end{center}
\end{figure}

According to Eq.~(\ref{eq:1loop_g2}), $\Delta a^{1,H/A^0}_\mu$ and  $\Delta a^{1,H^\pm}_\mu$ with $c_{\beta-\alpha}=0$ depend only on the scalar boson masses and the parameter $\zeta_\ell$.  Thus, we show $\Delta a^{1,X}_\mu$  as a function of $m_X$ for $X=H,A^0, H^\pm$ in Fig.~\ref{fig:1loop_g2}(a), (b), and (c), respectively, where the curves correspond to  $|\zeta_\ell|=(30,50,100)$.  Clearly, the pseudoscalar $A^0$ and the charged-Higgs $H^\pm$ contributions are always negative at the one-loop level.  The $H^\pm$ contribution is small and can be neglected.  Although $H$ makes a positive contribution to $\Delta a_\mu$, $\Delta a^{1,H}_\mu > 10^{-9}$ is possible only when $m_H$ is lighter than about 50~GeV.  According to the global fit results in the A2HDS with $m_H< m_h$ scenario~\cite{Eberhardt:2020dat}, such light CP-even scalar is still allowed.  However, the associated $m_{A^0}$ parameter can be of $O(100)$~GeV.  To demonstrate the correlation between $m_H$ and $m_{A^0}$, the contours of the combined $\Delta a_\mu$ as a function of $m_{H}$ and $m_{A^0}$ are shown in Fig.~\ref{fig:1loop_g2}(d), where $\zeta_\ell=-100$ and $m_{H^\pm}=180$~GeV are used. When $A^0$ with the allowed mass of $O(10^2)$~GeV is included, comparing to the case without $A^0$ contribution, the $m_H$ value required for $\Delta a_\mu > 10^{-9}$ has to be shifted downward.  Hence, in the region of $m_H > 100$~GeV, the one-loop contribution to the muon $g-2$ within the 2HDM is far below $10^{-9}$.

\subsubsection{Barr-Zee contribution within the 2HDM}

\begin{figure}[phtb]
\begin{center}
\includegraphics[scale=0.63]{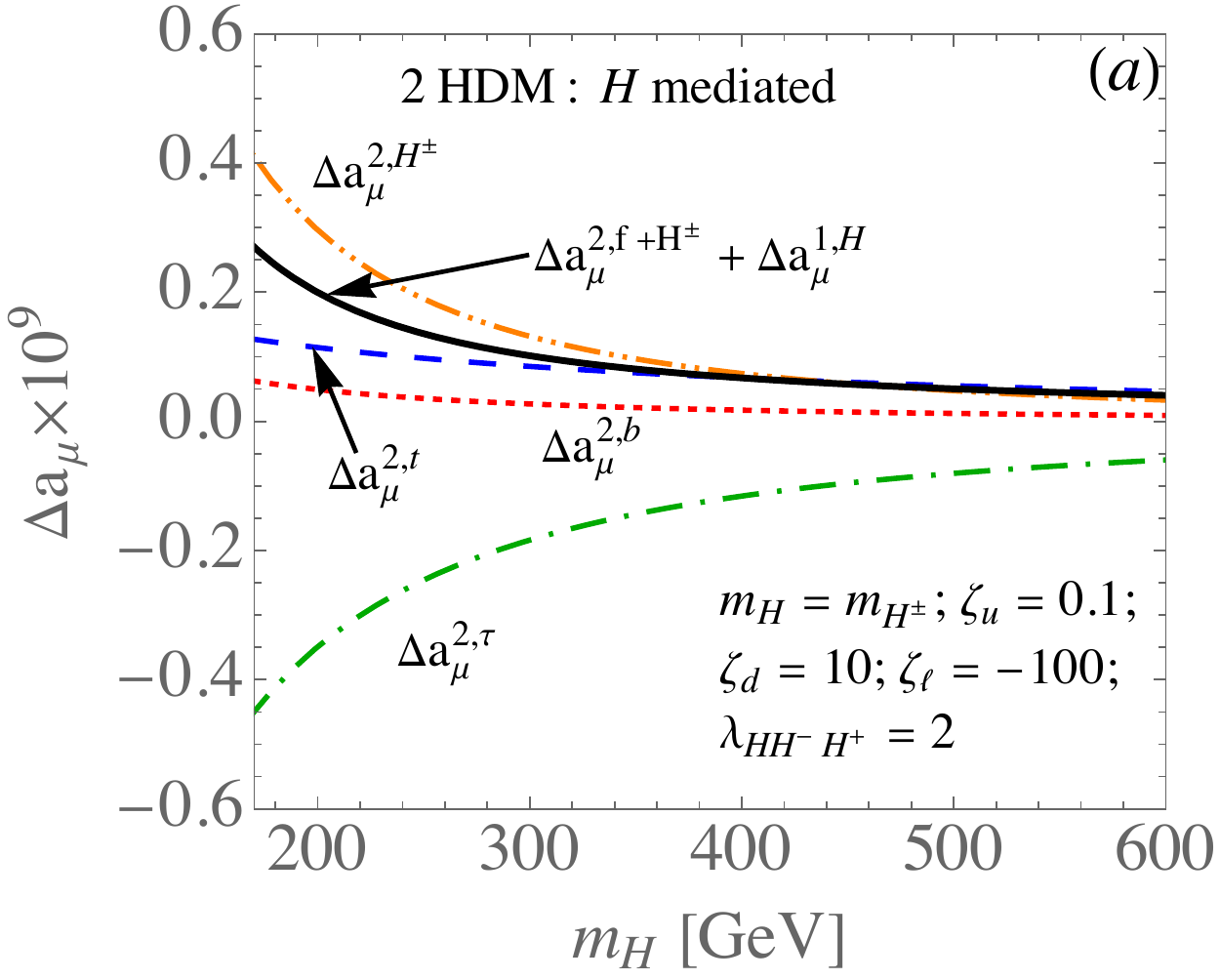}
\hspace{5mm}
\includegraphics[scale=0.58]{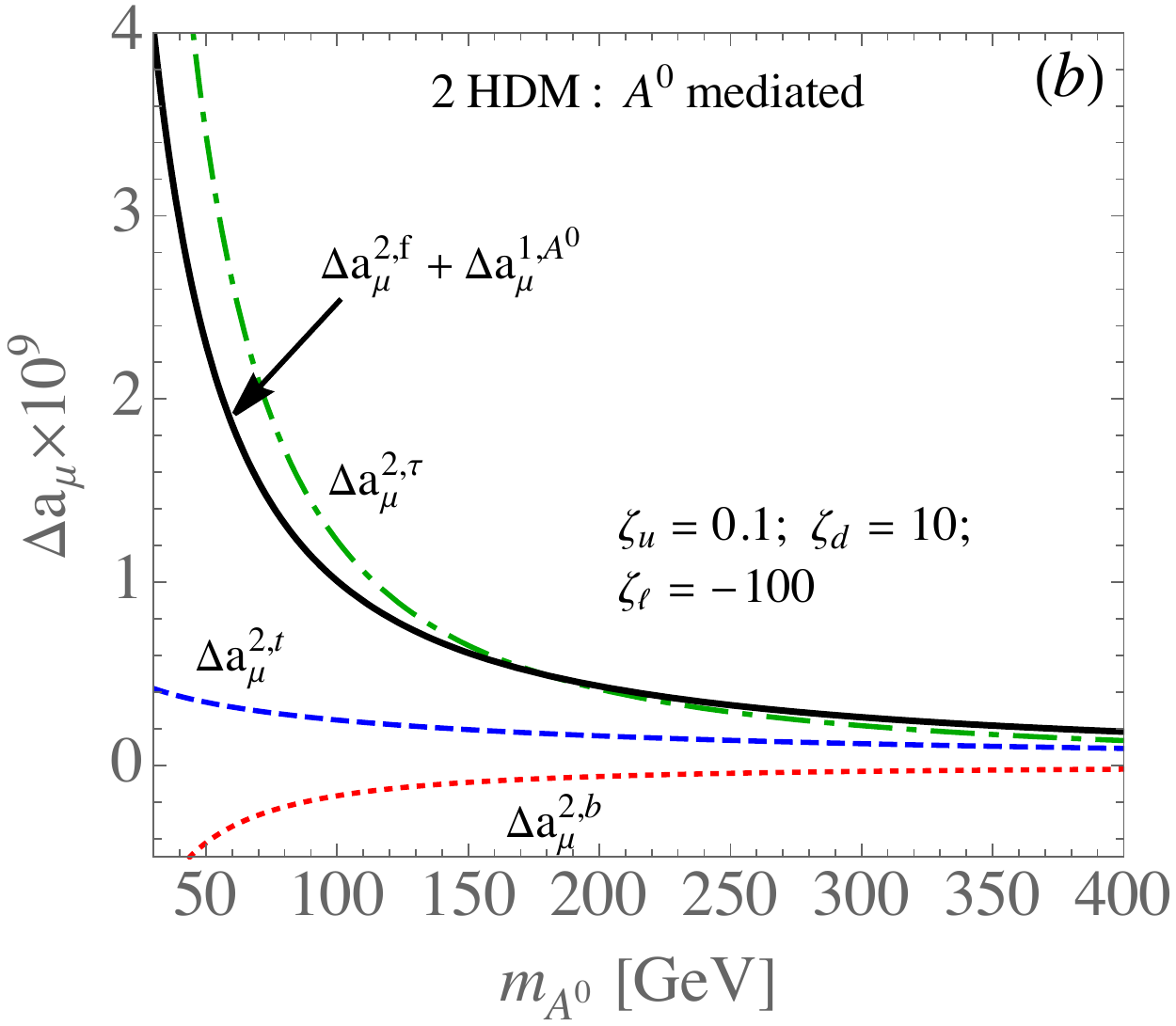}
 \caption{ Two-loop muon $g-2$ from Figs.~\ref{fig:1_2loop_Feyn}(c) and~\ref{fig:1_2loop_Feyn}(d)  induced by the mediation of  (a) $H$ and  (b) $A^0$ as a function of respectively $m_{H, A^0}$, where the dashed, dotted, dot-dashed, and dot-dot-dashed curves are the top, bottom, $\tau$-lepton, and $H^\pm$ contributions, respectively.  The solid curves contain both one-loop and two-loop effects.  }
\label{fig:2loop_2HDM}
\end{center}
\end{figure}

Before discussing the new two-loop effects on $\Delta a_\mu$ in our model, we first focus on the analysis within the 2HDM. 
As discussed before, the main two-loop Barr-Zee diagram contributions are usually from the fermion and charged Higgs loops shown in Figs.~\ref{fig:1_2loop_Feyn}(c) and \ref{fig:1_2loop_Feyn}(d), respectively.  In order to understand the influence of these effects on $\Delta a_\mu$, we separately show the fermion and $H^\pm$ contributions in Fig.~\ref{fig:2loop_2HDM}, where plot (a) [(b)] is mediated by $H~[A^0]$; the dashed, dotted, dot-dashed, and dot-dot-dashed curves are the top, bottom, $\tau$-lepton, and $H^\pm$ contributions, respectively, and the solid curves are the combined fermion- and $H^\pm$-loop Barr-Zee contributions and the one-loop results.  In the numerical estimates, we take $\zeta_u=0.1$, $\zeta_b=10$, and $\zeta_\ell=-100$, consistent with the numerical results given in Ref.~\cite{Eberhardt:2020dat}.  To estimate the $H^\pm$ contribution, we use the conservative value of $\lambda_{HH^-H^+}=2$ and set $m_{H^\pm}=m_H$.  In the $H$-mediated part given in Fig.~\ref{fig:2loop_2HDM}(a), it can be seen that due to the strict bounds on $\zeta_{u,d}$ the top- and bottom-quark contributions are smaller than $2\times 10^{-10}$, although an enhancement factor of $\zeta_\ell$ is already applied.  The $H$-mediated $\tau$-loop contribution is always negative and sizable in magnitude.  Intriguingly, using $\lambda_{HH^-H^+}=2$, it is found that the $H^\pm$-loop in the mass region of $m_{H,H^\pm} > m_h$ gives the dominant effect, and it can overcome the negative $\tau$-loop contribution, so that $\Delta a_\mu$ is positive.

In the $A^0$-mediated part given in Fig.~\ref{fig:2loop_2HDM}(b), similar to the case mediated by $H$, the top-quark and bottom-quark effects are small.  However, the bottom-quark and $\tau$-lepton contributions interchange sign, and the latter becomes the dominant effect.  The sign difference arises from the loop functions $J^H_\tau(z)$ and $J^{A^0}_\tau(z)$ shown in Eq.~(\ref{eq:loop_fun}).  For the region of $m_{A^0}>m_h$, it can be seen from Fig.~\ref{fig:2loop_2HDM}(b) that its two-loop effect on $\Delta a_\mu$ is smaller than $10^{-9}$.  Nevertheless, when  $A^0$ is lighter than the SM-like Higgs, its contribution to $\Delta a_\mu$ increases significantly.  When the negative one-loop contribution is included, one observes that the $A^0$ contribution to $\Delta a_\mu$ can reach $2\times 10^{-9}$ when $m_{A^0}\approx 55$~GeV.  Following the global analysis in the A2HDS~\cite{Eberhardt:2020dat}, when $m_H>m_h$, such a light pseudoscalar boson is not excluded by the current experimental data.  Hence, the conclusion is consistent with that obtained in the 2HDM type-X~\cite{Broggio:2014mna,Han:2015yys}.

\subsubsection{Barr-Zee contribution from the triplet extension}

It has been shown that when the CP-even and CP-odd scalar masses are heavier than the observed Higgs mass, the $H$-mediated and $A^0$-mediated effects in the 2HDM become ineffective to accommodate the measured $\Delta a_\mu$.  In the following analysis, we discuss the new contributions from the doubly- and singly-charged Higgs bosons derived from the Higgs triplet extension.  In the analysis, we focus on the $m_H > m_h$ scenario, following the parameter constraints given in Ref.~\cite{Eberhardt:2020dat}.

\begin{figure}[phtb]
\begin{center}
\includegraphics[scale=0.6]{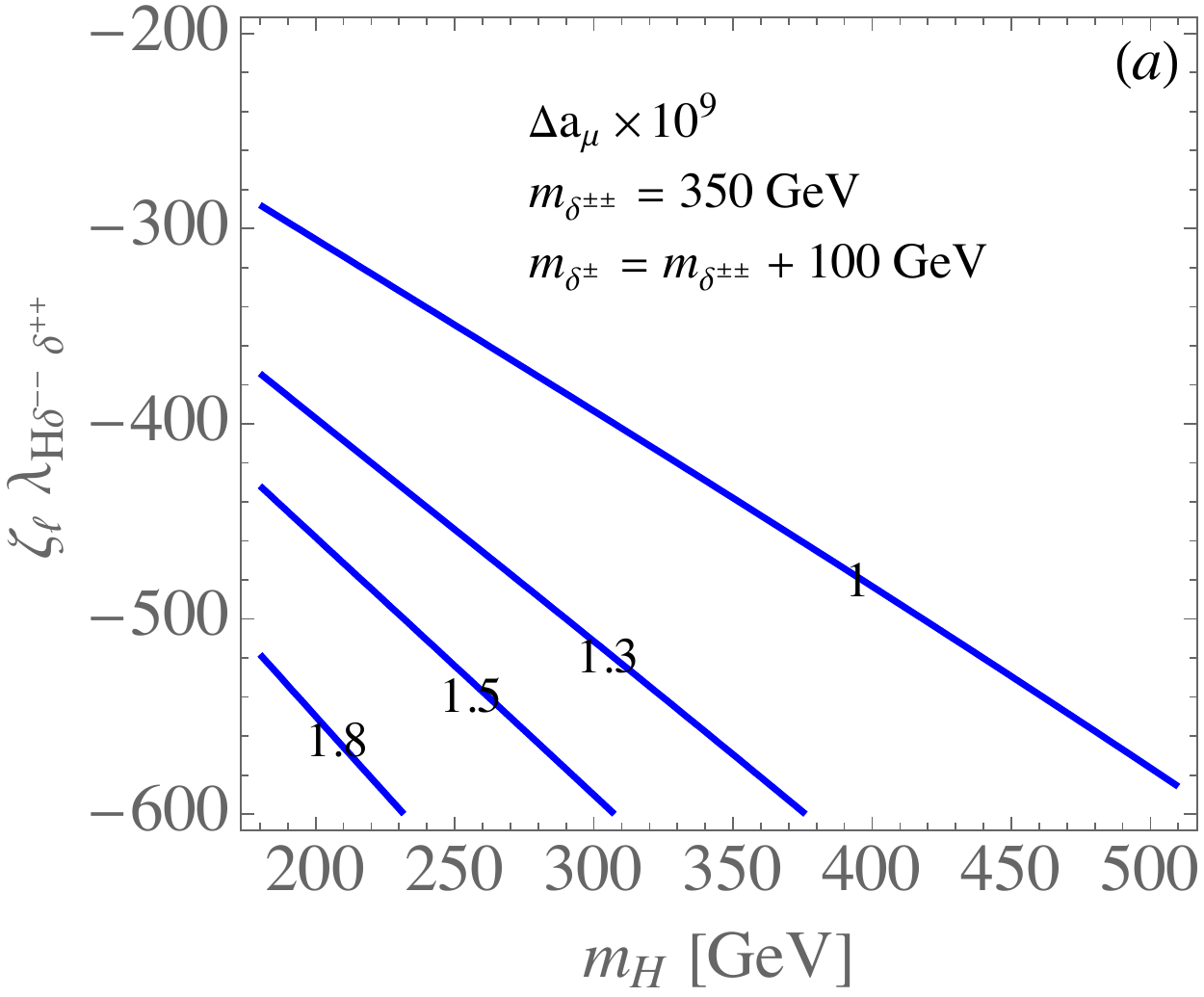}
\hspace{5mm}
\includegraphics[scale=0.6]{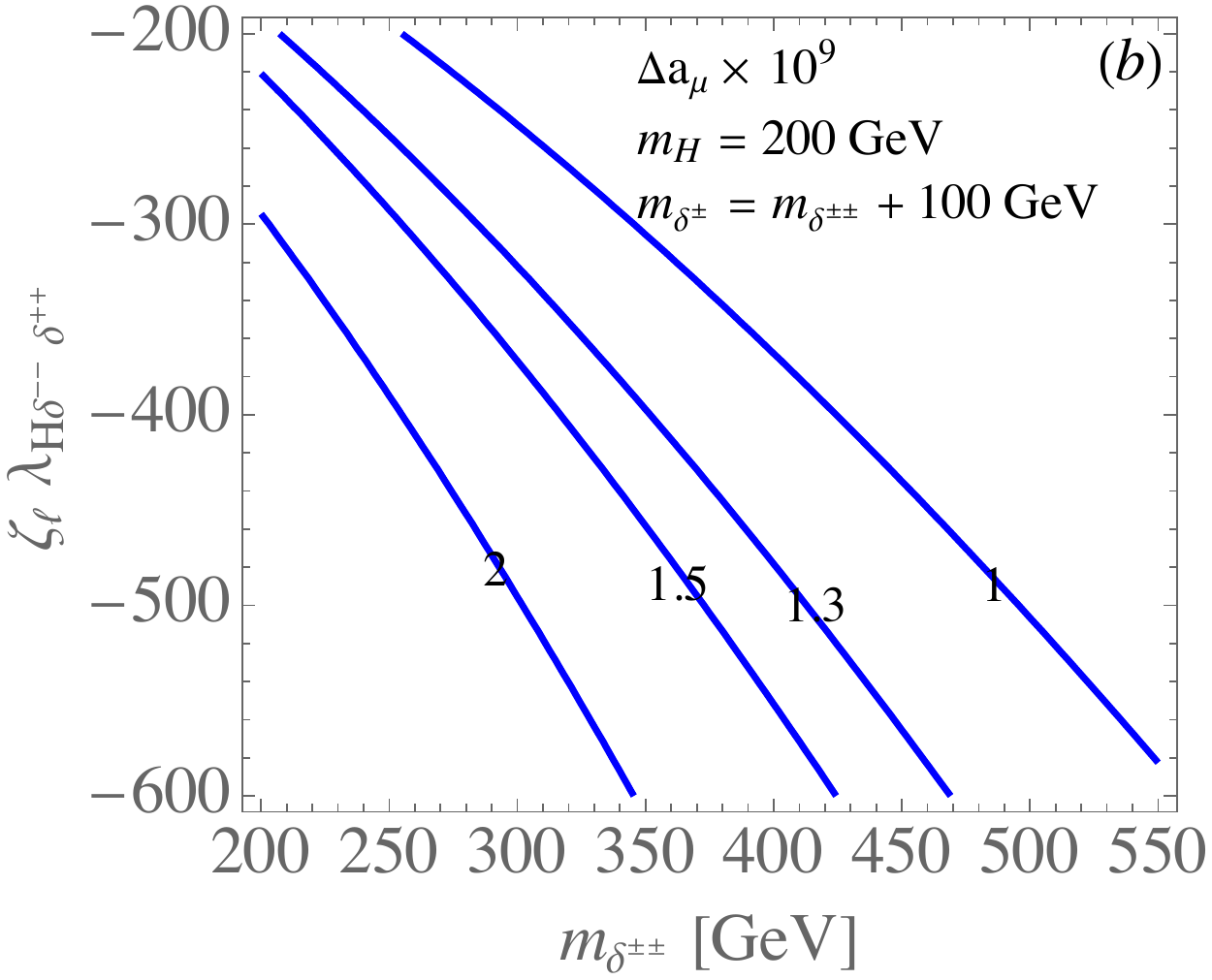}
 \caption{ Contours of $\Delta a_\mu$ from the 2-loop Barr-Zee type diagrams of $\delta^{\pm\pm}$ and $\delta^{\pm}$ in the plane of (a) $m_H$ and $\zeta_\ell \lambda_{H\delta^{--}\delta^{++}}$ with $m_{\delta^{\pm\pm}}=350$~GeV and (b) $m_{\delta^{\pm\pm}}$ and $\zeta_\ell \lambda_{H\delta^{--}\delta^{++}}$ with $m_H=200$~GeV. }
\label{fig:2loopHD}
\end{center}
\end{figure}

According to Eq.~(\ref{eq:2loop_g2}), in addition to the mass factor $\Delta a^{2,Y}_\mu$ ($Y=\delta^{++}, \delta^{+}$) further depends on the product $\zeta_\ell \lambda_{HY^* Y}$. Since the $\delta^+$-loop effect is similar to the $H^+$-loop effect, its contribution is expected to be of ${\cal O}(10^{-10})$.  However, the doubled electric charge of $\delta^{++}$ results in a factor of 4 enhancement.  We show contours of $\Delta a^{2,\delta^{++} + \delta^+}_\mu$ in the $m_H$-$\zeta_\ell \lambda_{H\delta^{--} \delta^{++}}$ plane in Fig.~\ref{fig:2loopHD}(a), where  $\lambda_{H\delta^-\delta^+}=\lambda_{H\delta^{--}\delta^{++}}$, $m_{\delta^{\pm}}=m_{\delta^{\pm\pm}}+100$~GeV, and $m_{\delta^{\pm\pm}}=350$~GeV.  Fixing $\zeta_\ell=-100$, we treat $\lambda_{HY^*Y}$ as a variable and set $ \lambda_{HY^*Y}\lesssim 5$ to satisfy the perturbativity bound~\cite{Ilisie:2014hea}.  We observe that $\Delta a^{2,\delta^{++} + \delta^+}_\mu \approx 1.3\times 10^{-9}$ can be achieved in the model when $m_H\approx 200$~GeV and $\zeta_\ell \lambda_{H\delta^{--} \delta^{++}}\approx -320$.  Even when using the maximal value of $|\zeta_\ell \lambda_{H\delta^{--} \lambda^{++}}| = 500$, $\Delta a^{2,\delta^{++} + \delta^+}_\mu$ can still reach $10^{-9}$ at $m_{H}\approx 500$~GeV.  In Fig.~\ref{fig:2loopHD}(b), we show contours of $\Delta a^{2,\delta^{++} + \delta^+}_\mu$ in the $m_{\delta^{\pm\pm}}$-$\zeta_\ell \lambda_{H\delta^{--}\delta^{++}}$ plane, where  $m_H=200$~GeV is taken.  Clearly, with $\zeta_\ell \lambda_{H\delta^{--}\delta^{++}} \approx -500$, we can have $\Delta a^{2,\delta^{++} + \delta^+}_\mu\approx 10^{-9}$ at $m_{\delta^{\pm\pm}}\approx 500$~GeV.  These results demonstrate that the measured muon $g-2$ can be readily achieved even when the exotic Higgs bosons in our model have mass of a few $\times 100$~GeV.

\begin{figure}[phtb]
\begin{center}
\includegraphics[scale=0.7]{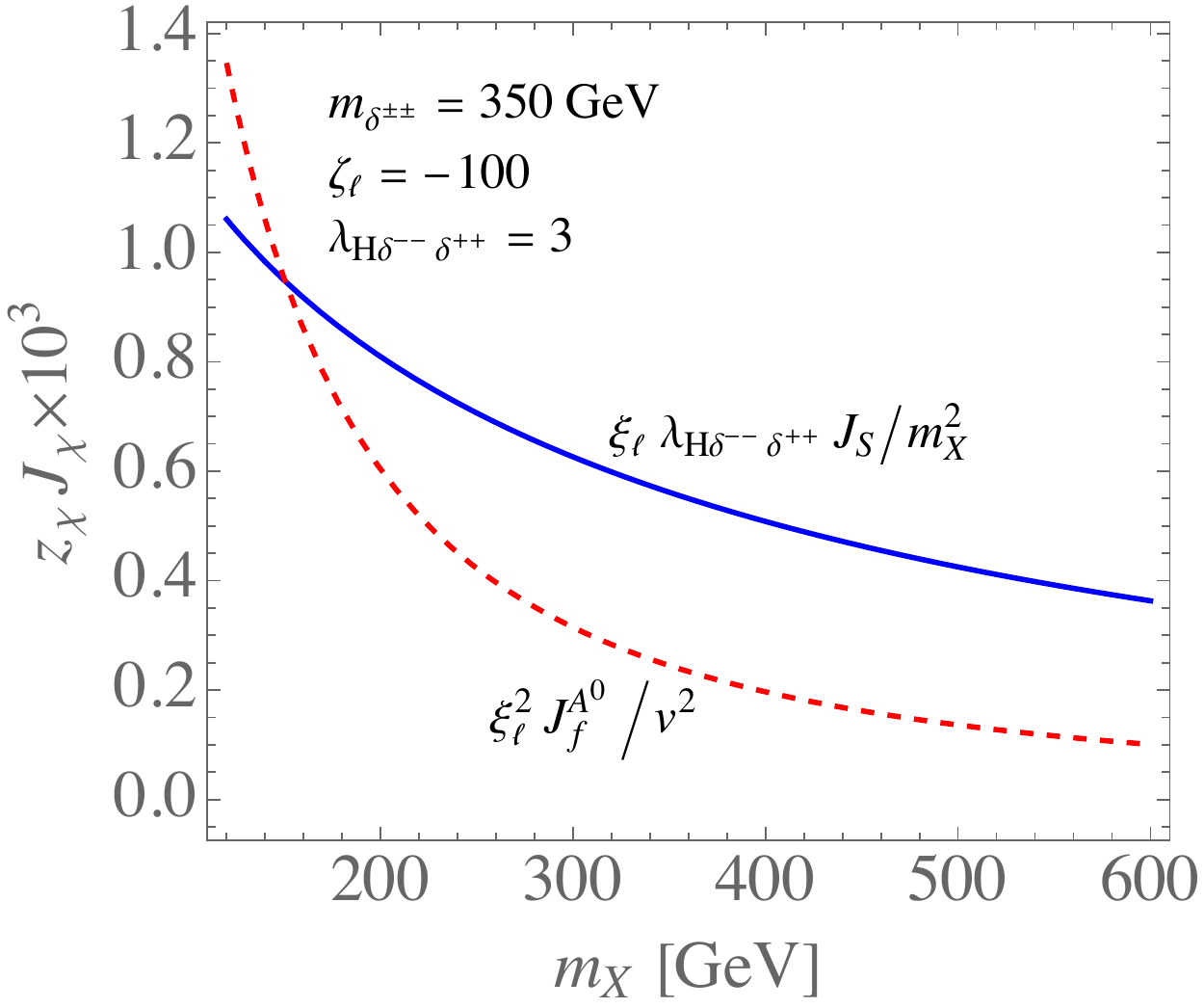}
 \caption{ Product of coupling factor and loop-integral function for the $\delta^{++}$-loop and $\tau$-loop.    }
\label{fig:loopInt}
\end{center}
\end{figure}

Besides the enhancement from the two units of electric charge, it is interesting to note the other enhancement factor associated with the doubly-charged Higgs boson by comparing the result with that induced from the $\tau$-loop.  Because the $H$ and $A^0$ couplings to muon are the same in the alignment limit, if we further set $m_H=m_{A^0}=m_X$ the only different factors come from the couplings, $A^0\tau\tau$ and $HH^-H^+$, and from the loop integrals, $J^{A^0}_f(z)$ and $J_S(z)$ defined in Eq.~(\ref{eq:loop_fun}).  For simplicity, we use $z_\chi J_\chi$ to represent the effect for the $\tau$-loop and the $\delta^{++}$-loop, where $z_\chi$ and $J_\chi$ are the associated coupling factor and integral function, respectively.  We show $z_\chi J_\chi$ as a function $m_X$ in Fig.~\ref{fig:loopInt}, where $\zeta_\ell=-100$, $m_{\delta^{++}}=350$~GeV, and $\lambda_{H\delta^{--} \delta^{++}}=3$ are applied.  It can be seen that once $m_X> 160$~GeV, the $\delta^{++}$-loop contribution is larger than the $\tau$-loop.

\begin{figure}[phtb]
\begin{center}
\includegraphics[scale=0.6]{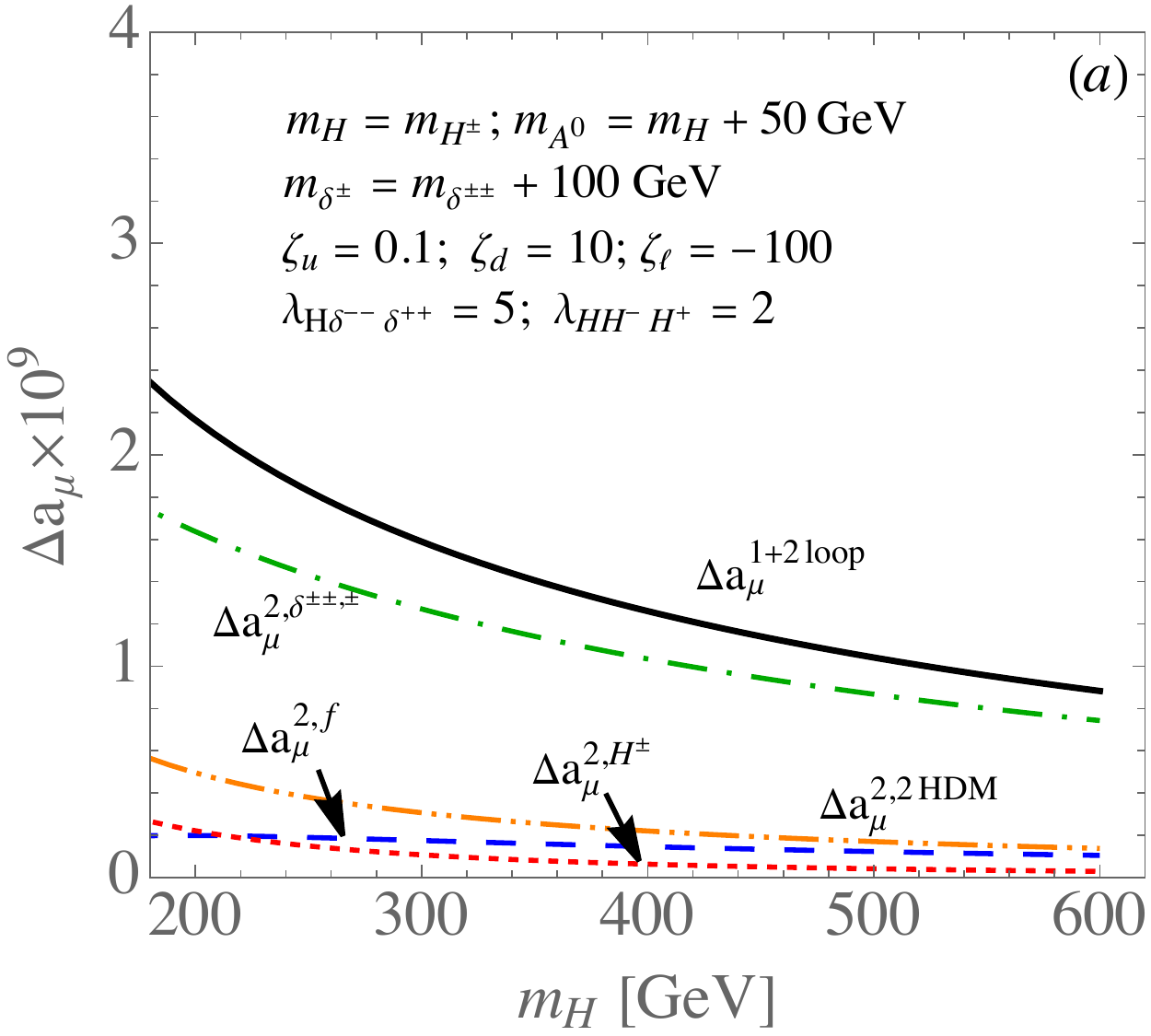}
\hspace{5mm}
\includegraphics[scale=0.6]{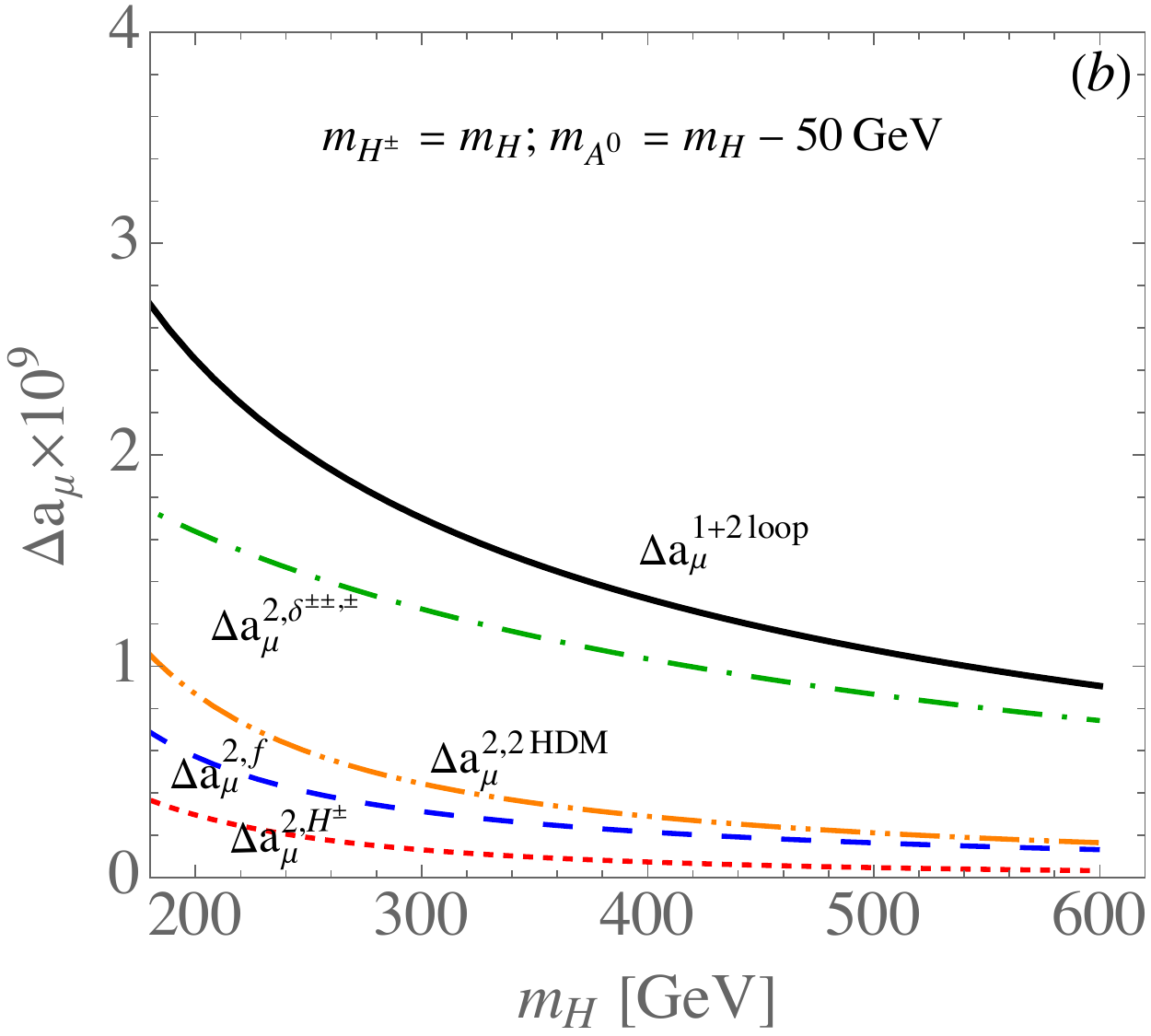}
 \caption{ Combined results of one-loop and two-loop contributions in the model for (a) $m_{H^\pm}=m_H$ and $m_{A^0}=m_H+50$~GeV and for (b) $m_{H^\pm}=m_H$ and $m_{A^0}=m_H-50$~GeV, where $m_{\delta^{\pm\pm}}=350$~GeV is used and the other parameters are the same as in plot (a).   }
\label{fig:1_2loop_total}
\end{center}
\end{figure}

\begin{figure}[phtb]
\begin{center}
\includegraphics[scale=0.7]{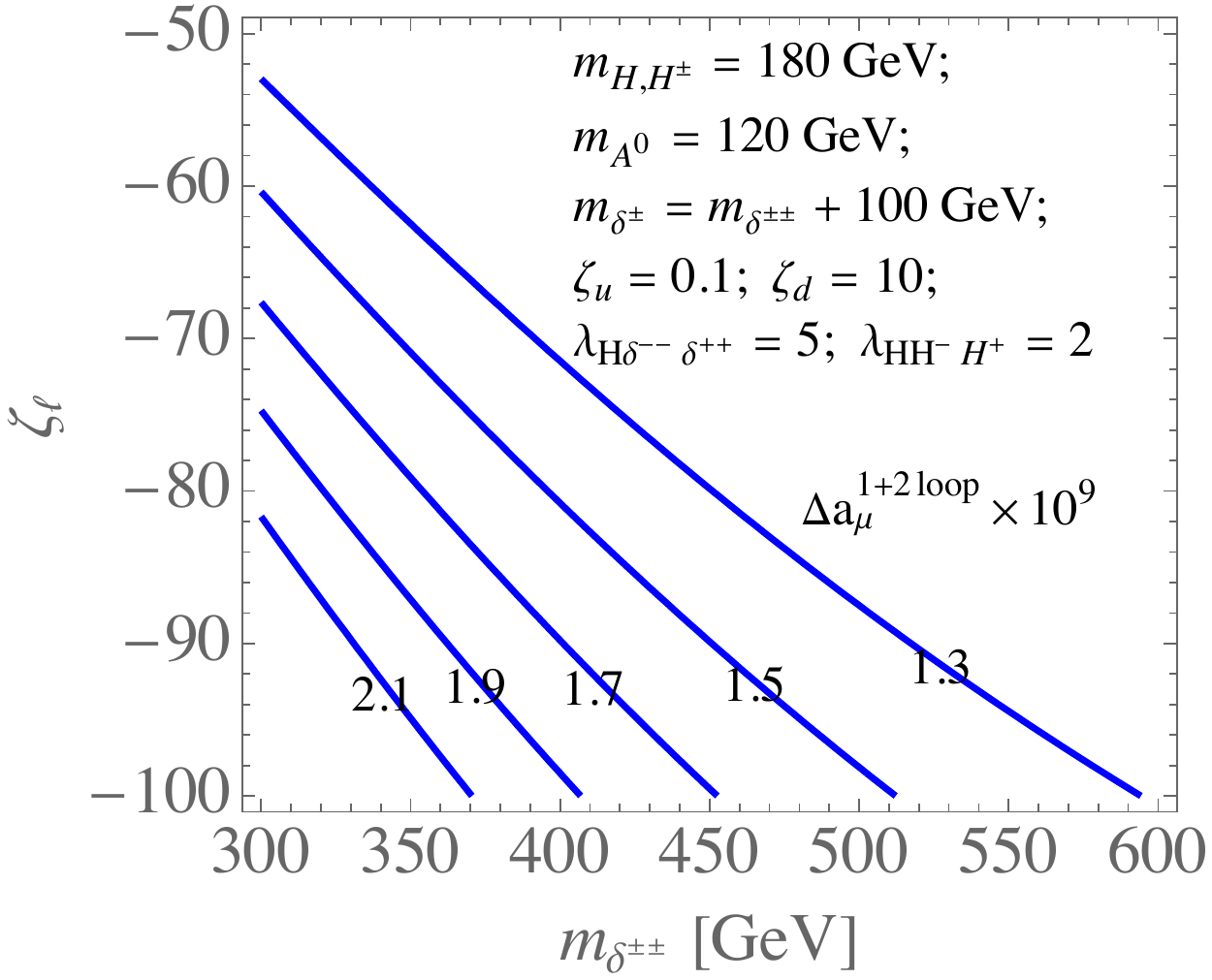}
 \caption{ Contours for the combined one- and two-loop contributions to $\Delta a_\mu$ (in units of $10^{-9})$ in the $\zeta_\ell$-$m_{\delta^{\pm\pm}}$ plane.   }
\label{fig:total}
\end{center}
\end{figure}

So far, we have just analyzed the effects of the Higgs triplet field on the muon $g-2$. Although the 2HDM contribution to $\Delta a_\mu$ becomes less significant as $m_{A^0}> m_h$, its effects are part of the model and we should combine them with the Higgs triplet effects altogether. According to the global fit analysis presented in Ref.~\cite{Eberhardt:2020dat}, the masses of $H$, $A^0$, and $H^\pm$ are strongly correlated and, in the $m_H>m_h$ scenario, $A^0$ and $H^\pm$ can be heavier or lighter than $h$.  For the purpose of illustration, we show the cases with $m_{H^\pm}=m_{H}$ for  $m_{A^0}=m_H+50$~GeV and for $m_{A^0}=m_H-50$~GeV in Fig.~\ref{fig:1_2loop_total}(a) and Fig.~\ref{fig:1_2loop_total}(b), respectively.  The dashed, dotted, and dot-dashed curves give respectively the fermion-loop, $H^\pm$, and $\delta^{++,+}$ contributions.  The dot-dot-dashed curves are the two-loop 2HDM results, where the fermion- and $H^\pm$-loop contributions are summed up.  The solid curves combine all the above-mentioned contributions, including the one-loop effects.  To show the maximal contribution from the Higgs triplet field, we take $\lambda_{H\delta^{--}\delta^{++}} = 5$ in the plots, and the other parameter are taken to be the same as those used in the earlier plots.  { In order to see the effects of $\zeta_\ell$ and $m_{\delta^{\pm\pm}}$ on $\Delta a_\mu$, we show the contours of $\Delta a_\mu$ (in units of $10^{-9}$) in the plane of $\zeta_\ell$ and $m_{\delta^{\pm\pm}}$ in Fig.~\ref{fig:total}, where the parameter values are taken to be the same as those used in Fig.~\ref{fig:1_2loop_total} with the exception of  $m_{H,H^\pm}=180$~GeV and $m_{A^0}=120$~GeV.  }

Based on the analyses, we summarize the results as follows:
\begin{enumerate}
\item 
{ The contribution of Barr-Zee type diagram with the $H^\pm$-loop can only be of ${\cal O}(10^{-10})$.}

\item
When the $H$- and $A^0$-mediated two-loop diagrams are combined, $A^0$ with a mass of $70$~GeV can lead to $\Delta a_\mu \sim 2 \times 10^{-9}$, similar to the situation of 2HDM type-X~\cite{Broggio:2014mna}.  However, when $m_{A^0}>m_h$, the 2HDM contribution is below $10^{-9}$.

\item
The $H$-mediated $\delta^{++(+)}$-loop contribution is independent of $m_{A^0}$, and it can play an important role on $\Delta a_\mu$ in a wide range of $m_H$ and $m_{\delta^{\pm\pm}}$, particularly when the exotic Higgs masses are a few hundred GeV.
\end{enumerate}

Before closing this subsection, we make a brief remark about the implication of the model on the anomalous magnetic dipole moment of the electron (electron $g-2$).  Applying the accurate measurements of the fine structure constant from $^{133}$Cs and $^{87}$Rb to the theoretical calculations~\cite{Aoyama:2014sxa,ATN2019}, the differences in the electron $g-2$ between the experiments and the SM expectation are found to be:
\begin{align}
\Delta a_e (^{133}\text{Cs}) & = -(8.8\pm 3.6)\times 10^{-13}~~\text{\cite{PYZEM}}\,, \nonumber \\
\Delta a_e (^{87}\text{Rb}) &= (4.8 \pm 3.0)\times 10^{-13}~~\text{\cite{MYCG}}\,,
\end{align}
i.e., having $-2.4\sigma$ and $1.6\sigma$ deviations, respectively.
Their weighted average is $\Delta a_e = -0.8 \pm 2.3$.  In spite of the inconclusively experimental results, the model predicts a concrete correlation in the corrections for muon $g-2$ and electron $g-2$.  Since the dominant contribution comes from the Barr-Zee type diagrams in the model and the lepton Yukawa couplings to $H(A^0)$ are proportional to $m_\ell$ according to Eq.~(\ref{eq:2loop_g2}), we have the ratio $\Delta a_e /\Delta a_{\mu} = (m_e/m_\mu)^2\simeq 2.36\times 10^{-5}$.  Hence,  $\Delta a_e$ and $\Delta a_\mu$ should have the same sign in the model.

\subsection{Doubly-charged Higgs decays}

In this subsection we discuss the decay branching ratios of $\delta^{\pm \pm}$ in our scenario and its related collider signature.  With our choice of $v_\Delta > \mathcal{O}(10^{-4})$ GeV, $\delta^{\pm \pm}$ dominantly decays into charged scalar and/or $W$ bosons, evading the stringent constraint of LHC searches for same-sign dileptons.
The interactions relevant to the dominant decays of $\delta^{\pm \pm}$ are obtained from the scalar potential and kinetic term as follows
\begin{align}
\label{eq:int-decay}
\mathcal{L} \supset & -(\mu_1 s_\beta^2 + \mu_2 c_\beta^2 - \mu_3 s_\beta c_\beta) \delta^{++} H^- H^- - \frac{v}{2} \left( s_{2\beta} \frac{\lambda'_8-\lambda'_9} {2} - c_{2\beta} \lambda'_{12} \right) \delta^{++} H^- \delta^-   \nonumber\\
& - \frac{g^2}{\sqrt{2}} v_\Delta \delta^{++} W^{-\mu} W^-_\mu  - ig W^{-\mu} (\partial_\mu \delta^- \delta^{++} - \delta^- \partial_\mu \delta^{++}) + \mbox{h.c.} \ ,
\end{align}
where $g$ is the $SU(2)_L$ gauge coupling. 
It can be seen that the $\delta^{++} H^{-} H^{-}$ and $\delta^{++} W^- W^-$ couplings are dictated by the factor of $v_\Delta$. That is, for the doubly-charged-Higgs decays, the mixing of ${\cal O}(v_\Delta)$ among $G^\pm$, $H^\pm$, and $\delta^{\pm}$ has to be taken into account though its effects are small in the muon $g-2$.  If we take $\mu_3=t_{2\beta} (\mu_1-\mu_2)$ and $\lambda_8 c^2_\beta + \lambda_9 s^2_\beta + \lambda'_{12} s_{2\beta} \sim - 2 (2m_\Delta/v)^2$ in Eq.~(\ref{eq:Charged_M}), it is found that $m^2_{G^- \delta^+}\sim {\cal O}(v^2_\Delta)$ and  $m^2_{H^- \delta^+} \sim {\cal O}(v_\Delta)$.  With such parameters and dropping ${\cal O}(v^2_\Delta)$ terms, the relevant charged scalar mass-square matrix can be simplified as a $2\times 2$ matrix:
 \begin{equation}
  \begin{pmatrix} H^- \delta^- \end{pmatrix}
  \begin{pmatrix}
 m^2_{H^{\pm}} & m^2_{H^- \delta^+} \\ 
 m^2_{H^- \delta^+} & m^2_{\delta^\pm} \\ 
  \end{pmatrix}  \begin{pmatrix}
    H^+  \\ 
    \delta^+ \\ 
  \end{pmatrix}\,, \label{eq:CH2b2_m}
 \end{equation}
where the  matrix can be diagonalized by an $SO(2)$ transformation, similar to the expression given in Eq.~(\ref{eq:H-Phi}) but using the $\theta_\pm$ mixing angle instead of $\alpha$.  Due to the fact that $v_\Delta \ll v$, we can ignore the influence of $m^2_{H^- \delta^+}$ on $m^2_{H^\pm}$ and $m^2_{\delta^\pm}$, and the physical states and the small mixing angle can be simply expressed as:
 \begin{align}
 \begin{split}
 H^\pm_1 & \simeq  H^\pm +  \theta_\pm \, \delta^{\pm}
 ~,
 \\
 H^\pm _2 & \simeq -\theta_\pm \, H^\pm + \delta^\pm
 ~, 
 \\
\theta_{\pm} & \simeq \frac{v v_\Delta }{2\sqrt{2}(m^2_{\delta^\pm} - m^2_{H^\pm})}  \left( s_{2\beta} \frac{\lambda'_8 - \lambda'_9}{2} + s^2_\beta \lambda'_{12}\right)
~.
 \end{split}
 \end{align}
Taking $\lambda'_8-\lambda'_9 =1$, $\lambda'_{12}=1$, $m_{\delta^\pm}=350$~GeV, $m_{H^\pm}=180$~GeV, $v_\Delta =5\times 10^{-4}$ and $c_\beta=s_\beta=1/\sqrt{2}$, the mixing angle value is estimated to be $\theta_\pm\simeq 4.8\times 10^{-7}$.
Since $H^\pm_{1(2)}$ only carries a tiny component of $\delta^\pm (H^\pm)$, in the following analysis we keep using $H^\pm(\delta^\pm)$ instead of $H^\pm_1(H^\pm_2)$.

Because of the introduction of $\theta_\pm$, in addition to the $W^\pm W^\pm$ and $H^\pm H^\pm$ modes, the doubly-charged Higgs can also decay into $H^\pm W^\pm$.  In the case an on-shell decay to $H^\pm H^\pm$ is impossible, we consider the decay channel $H^\pm H^{\pm *}$, {\it i.e.}, one of $H^\pm$ is off-shell, with the assumption that the $H^\pm \to \tau^\pm \nu$ is dominant.  The partial decay widths of these dominant modes are explicitly expressed as:
\begin{align}
\Gamma(\delta^{\pm \pm} \to W^\pm W^\pm) 
\simeq & 
\frac{g^4 v_\Delta^2 m_{\delta^{\pm \pm}}^3}{16 \pi m_{W}^4} \left( 3 r_W^2 - r_W + \frac14  \right) \sqrt{1-4r_W} ~, \\
\Gamma(\delta^{\pm \pm} \to H^\pm W^\pm) 
\simeq &
\frac{g^2 m^3_{\delta^{\pm \pm}}}{16 \pi m_W^2} \theta_\pm^2  \lambda(1,r_W, r_{\pm})^{\frac23} ~,  \\
\Gamma(\delta^{\pm \pm} \to H^\pm H^\pm)
\simeq &
\frac{v^2 \sqrt{1 - 4 r_\pm}}{4 \pi m_{\delta^{\pm \pm}}} |\lambda_{\delta^{++} H^- H^-}|^2 ~, \\
\Gamma(\delta^{\pm\pm} \to H^{\pm} H^{\pm*}) 
\simeq &
\frac{1}{2^6 \pi^3 m_{\delta^{\pm\pm}}} \left| \sqrt{2} m_\tau \zeta_\ell \lambda_{\delta^{++} H^- H^-} \right|^2  \nonumber \\
& \times \int^{x_{\rm max}}_{x_{\rm min}} dx  \frac{x}{(x -r_\pm)^2} \sqrt{ \lambda\left(1, r_{\pm}, x \right)}~,
\end{align}
where $r_W=m^2_W/m^2_{\delta^{\pm\pm}}$, $r_\pm = m_{H^\pm}^2/m_{\delta^{\pm \pm}}^2$; $x=q^2/m^2_{\delta^{\pm\pm}}$ and  $q^2$ is the invariant mass of $\tau \nu_\tau$ in the $H^+ \to \tau^+ \nu_\tau$ decay; $x_{\rm max}= (1-  \sqrt{r_\pm})^2$, $x_{\rm min} = m^2_\tau/m^2_{\delta^{\pm\pm}}$;  $\lambda(1,x,y) = 1 + x^2 +y^2  -2xy -2 x - 2y$, and 
 \begin{equation}
\lambda_{\delta^{++} H^- H^-} = \frac{1}{v}\left(\mu_1 s_\beta^2 + \mu_2 c_\beta^2 - \mu_3 s_\beta c_\beta \right)+ \frac{ 1}{2} \left( s_{2\beta} \frac{ \lambda'_8-\lambda'_9}{2} - c_{2\beta} \lambda'_{12} \right) \theta_\pm \,.
 \end{equation} 
 { For the purpose of comparison, we include the  $\delta^{++} \to \ell^+_i \ell^+_j$ channels. According to the $\delta^{++}$ Yukawa couplings to $\ell^-_i \ell^-_j$ shown in Eq.~(\ref{eq:yu_neutrino}), the decay rate is given by:
  \begin{equation}
  \Gamma(\delta^{++} \to \ell^+_i \ell^+_j )= \frac{m_{\delta^{\pm\pm}}}{8\pi (1+ \delta_{ij})} \left| \frac {(M_\nu)_{ij}}{v_\Delta}\right|^2\,, \label{eq:BRll}
  \end{equation}
where $\delta_{ij}$ is the Kronecker delta. We use the  $(M_\nu)_{ij}$ values determined by a global fit to the neutrino data and given in~\cite{deSalas:2017kay}.
}

\begin{figure}[phtb]
\begin{center}
\includegraphics[scale=0.6]{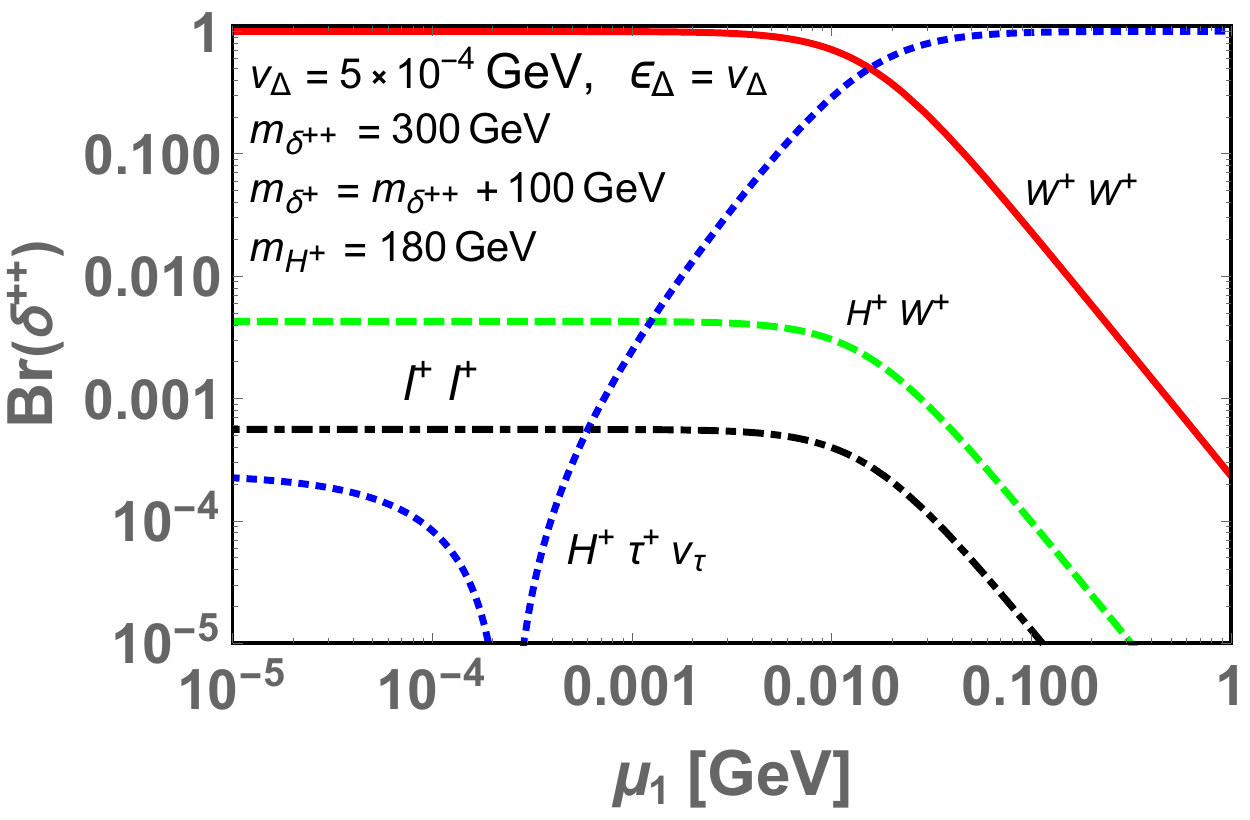}
\hspace{5mm}
\includegraphics[scale=0.6]{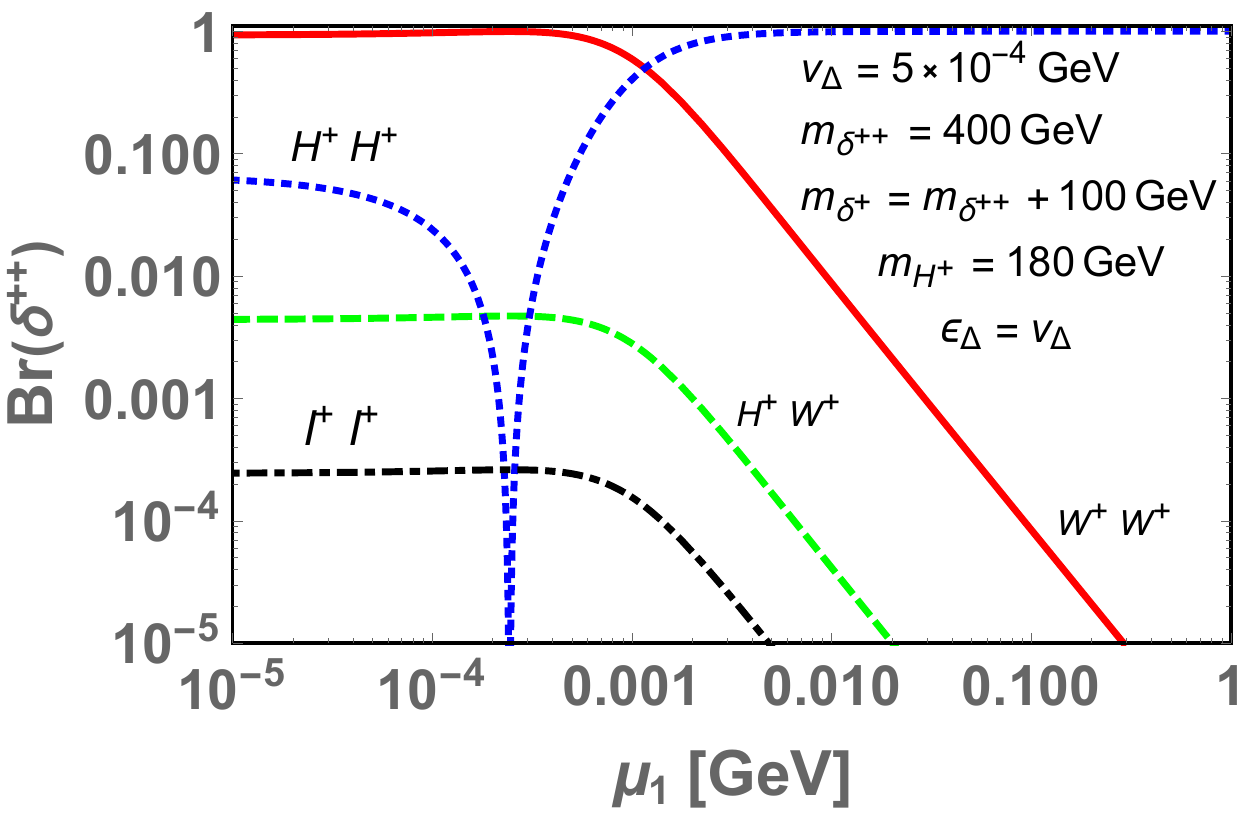}
 \caption{ Branching ratios of $\delta^{++}$ as functions of $\mu_1$, with $t_\beta = 1$, $\epsilon_\Delta = v_\Delta = 5 \times 10^{-4}$~GeV, $m_{\delta^{\pm\pm}}= 300~(400)$~GeV and $m_{H^\pm} = 180$~GeV for left (right) plots.    }
\label{fig:BRdelta}
\end{center}
\end{figure}

\begin{figure}[phtb]
\begin{center}
\includegraphics[scale=0.8]{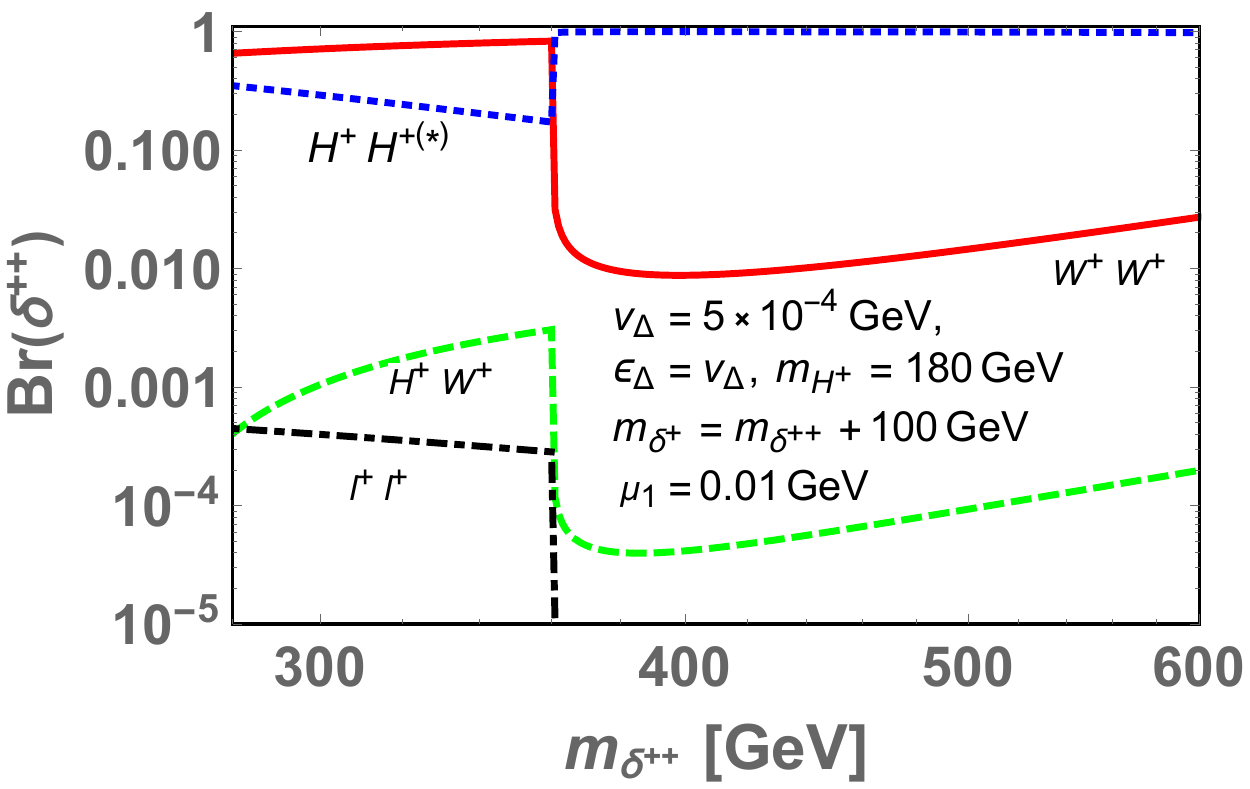}
 \caption{ Branching ratios of $\delta^{++}$ as functions of $m_{\delta^{\pm \pm}}$, with $t_\beta = 1$, $\epsilon_\Delta = v_\Delta = 5 \times 10^{-4}$~GeV, $\mu_1=0.01$~{\rm Ge}V and $m_{H^\pm} = 180$~GeV.    }
\label{fig:BRdelta2}
\end{center}
\end{figure}

For illustration purposes, we show in Fig.~\ref{fig:BRdelta} their branching ratios { as functions of $\mu_1$} with the parameter choice of $t_\beta = 1$, $|\zeta_\ell| =100$, $\epsilon_\Delta = v_\Delta = 5 \times 10^{-4}$~GeV, $m_{\delta^{\pm\pm}}= 300~(400)$~GeV and $m_{H^\pm} = 180$~GeV for the left (right) plot  where Eq.~(A3) is applied to fix $\mu_{2}$ and $\mu_3$. 
{ For the neutrino mass matrix element in Eq.~\eqref{eq:BRll}, we apply the dominant ones $(M_\nu)_{\mu \mu,\tau\tau,\mu\tau} \sim 2 \times 10^{-2}$~eV according to the global fit in~\cite{deSalas:2017kay} for the normal ordering of neutrino masses.}
Note that one of $H^+$ is off-shell in the left plot while both $H^+$ are on-shell in the right plot.
In the former case, we find that the $H^+H^{+*}(\to \tau^+ \nu)$ mode can be dominant when $\mu_1 \gg v_\Delta$, even though it is a three-body decay.
In the latter case, we find that the $H^+H^+$ mode is dominant for $\mu_1 > \mathcal{O}(v_\Delta)$ and that  the ratio of branching ratio is ${\rm Br}(H^+H^+):{\rm Br}(W^+W^+) \simeq 0.07:1$ when $\mu_1 \lesssim v_\Delta$ except for the region $\mu_1 \sim v_\Delta/2$.  
 The suppression of ${\rm Br}(H^+H^+)$ at around $\mu_1 = v_\Delta / 2$ is due to a cancellation in the coupling.  The cancellation occurs when Eq.~(A3) is applied.  With $t_\beta=1$,  we have $\mu_1=\mu_2$ and $\mu_3= 2(\epsilon_\Delta - \mu_1)$, $\mu_1 s^2_\beta + \mu_2 c^2_\beta - \mu_3 c_\beta s_\beta \simeq 2\mu_1 - \epsilon_\Delta$.  The mixing angle $\theta_\pm$ could induce a sizable effect if $\lambda'_8-\lambda'_9=2$ is used.
Note also that $\lambda'_{12}$ does not contribute to the coupling $\lambda_{\delta^{++ H^- H^-}}$ when $t_\beta =1$ is taken.
In Fig.~\ref{fig:BRdelta2}, we also show the branching ratios of $\delta^{++}$ as functions of $m_{\delta^{++}}$ with $\mu_1 = 0.01$~GeV, where the other parameter values are the same as those shown in Fig.~\ref{fig:BRdelta}.
We find that the $H^+H^{+(*)}$ mode becomes dominant when $m_{\delta^{\pm \pm}} > 2m_{H^+}$ and that the branching ratios of $W^+W^+$ and $H^+W^+$ slightly increase with $m_{\delta^{\pm \pm}}$.

Finally we discuss signals from doubly-charged Higgs boson production at the LHC.
The $\delta^{++}\delta^{--}$ pair can be produced via electroweak interactions in proton-proton collision process. 
{ In our scenario, the produced $\delta^{\pm \pm}$ dominantly decay into $W^\pm W^\pm$ and/or $H^\pm H^{\pm (*)}$ mode, depending on the value of $\mu_1$ as discussed above. 
Here we focus on the $H^\pm H^{\pm (*)}$ mode as it is a special channel in our model.  The scenario when the $W^\pm W^\pm$ mode is more dominant is equivalent to the Higgs triplet model with $v_\Delta > \mathcal{O}(10^{-4})$ GeV.  
The singly-charged Higgs $H^\pm$ dominantly decays into $\tau \nu$ mode since the Yukawa interaction between $H^\pm$ and leptons is enhanced by the large $\zeta_\ell$ factor that is required for a sizable muon $g-2$ contribution.     
Therefore, the signature from $\delta^{++}\delta^{--}$ production is $4 \tau + \slashed{E}_T$ in our scenario.
Note that we can relax the bound on $m_{\delta^{\pm\pm}}$ of $\sim 350$~GeV~\cite{Aad:2021lzu} because the analysis assumes that the $W^\pm W^\pm$ mode is dominant and considers only the muons/electrons in the final state.
It would be difficult to reconstruct the doubly-charged Higgs mass because of the missing transverse energy carried away by neutrinos from $H^\pm$ decays.
Our signature could be tested in multi-tau searches in future LHC experiments.
}


\section{Summary}\label{sec:summary}

In this work, we have studied an extension of the Standard Model only in the scalar sector, with the addition of one Higgs doublet and one complex Higgs triplet, rending a two-Higgs-doublet model (2HDM) with the type-II seesaw mechanism.  For the 2HDM part, we consider the aligned two-Higgs-doublet scheme (A2HDS) to avoid undesired flavor-changing neutral currents induced by the two Higgs doublet fields and to satisfy the current Higgs data constraints.  The Higgs triplet field obtains a small vacuum expectation value (VEV) induced by the electroweak symmetry breaking and gives Majorana mass to neutrinos through Yukawa couplings.

We have examined how the model can accommodate the measured muon $g-2$ deviation.  Simple 2HDMs usually require CP-even and -odd Higgs bosons ($H$ and $A^0$) to be sufficiently light (about a few $\times {\cal O}(10)$~GeV) and rely on the contributions of Barr-Zee type diagrams to account for the muon $g-2$ anomaly, $\Delta a_\mu$.  In our model, the Barr-Zee type diagrams get additional contributions from a large $H\delta^{++(+)}\delta^{--(-)}$ coupling, an enhanced coupling between charged leptons and the charged Higgs boson ($H^\pm$) in 2HDM, and the electric charges of the charged Higgs bosons ($\delta^\pm$ and $\delta^{\pm\pm}$) from the Higgs triplet field, independent of the mass of CP-odd Higgs boson.  In fact, the mass of the exotic Higgs bosons is allowed to have a wider range, up to a few hundred GeV.

Owing to the new interactions with the other charged Higgs and $W$ bosons, the doubly-charged Higgs boson presents a different decay pattern than the usual Higgs triplet model.  With the assumed Higgs triplet VEV, $v_\Delta \sim 5 \times 10^{-4}$~GeV, the doubly-charged Higgs boson may dominantly decay into like-sign charged Higgs bosons in the 2HDM rather than like-sign $W$ bosons, { when the magnitudes of the trilinear couplings $\mu_{1,2,3}$ are greater than $10^{-3}~(10^{-2})$~GeV for both (one of the) charged Higgs bosons being on-shell.}
 Therefore, pair productions of the doubly-charged Higgs bosons will lead to the signature of 4 $\tau$-leptons and missing energy at the LHC.

\section*{Acknowledgments}
This work is supported in part by the Ministry of Science and Technology (MOST) of Taiwan under Grant Nos.~MOST-108-2112-M-006-003-MY2 (CHC) and MOST-108-2112-M-002-005-MY3 (CWC).

\appendix

\section{Mass matrices for neutral and charged Higgs bosons} \label{app:mass_matrix}

In this appendix, we show the full mass matrices at tree level for scalar, pseudoscalar, and charged Higgs bosons. Since $A^0$ and $H^\pm$ are the physical states in the 2HDM, it is useful to show the scalar mass matrices in terms of the  $(G^0, A^0, \eta^0)$ and $(G^+, H^+, \delta^+)$ bases when the Higgs triplet field $\Delta$ is introduced. Since the CP-even $H^0_1$ and $H^0_2$ scalars mix, we show the mass matrix in the basis of the $(h, H)$ states, defined in Eq.~(\ref{eq:2HDM_mass_Basis}).

From the scalar potentials given in Eqs.~\eqref{eq:v2_a}$-$\eqref{eq:v2_c} and the Higgs basis in Eq.~(\ref{eq:Higgs_Basis}), the mass matrix for the CP-odd components $G^0$, $A^0$ and $\eta^0$ is given by
\begin{equation}
\frac{1}{2} \begin{pmatrix} G^0 ~ A^0 ~ \eta^0 \end{pmatrix}
\begin{pmatrix} 
m^2_{G^0 G^0} & m^2_{G^0A^0} & m_{G^0 \eta^0}^2 \\ 
m^2_{G^0 A^0} & m_{A^0 A^0}^2 & m_{A^0 \eta^0}^2 \\ 
m_{G^0 \eta^0}^2 & m_{A^0 \eta^0}^2 & m_{\eta^0 \eta^0}^2 
\end{pmatrix}
\begin{pmatrix} G^0 \\ A^0 \\ \eta^0 \end{pmatrix},
\label{M_CP_odd}
\end{equation}
where the mass matrix elements
\begin{align}
m^2_{G^0 G^0} =& \frac{3 v_\Delta}{\sqrt{2}}  \left( c^2_\beta \mu_1+ s^2_\beta \mu_2 + c_\beta s_\beta \mu_3 \right)\,, \nonumber \\
m^2_{G^0 A^0} =& -\sqrt{2}  v_\Delta  \left[ s_{2\beta} (\mu_1 - \mu_2)  -c_{2\beta} \mu_3 \right]  \,, \nonumber \\
m^2_{G^0 \eta^0} =& - \sqrt{2}  v \left( c^2_\beta \mu_1+ s^2_\beta \mu_2 + c_\beta s_\beta \mu_3 \right)\,, \nonumber \\
m_{A^0 A^0}^2 \equiv& m^2_{A^0}= \frac{m^2_{12}}{s_\beta c_\beta} - \lambda_5 v^2  -\frac{v^2 }{2} \left( \lambda_6 t^{-1}_\beta + \lambda_7 t_\beta\right)  + 2\sqrt{2} s^2_\beta v_\Delta \mu_1 \nonumber \\
 & + \frac{c^2_\beta v_\Delta \mu_3}{\sqrt{2} t_\beta} (1-t^2_\beta)^2
  - \frac{v^2_\Delta \bar\lambda_{12}}{2} (s_{2\beta} + c^3_\beta + s^3_\beta )\,, \nonumber\\ 
m_{A^0 \eta^0}^2 =& \frac{v}{\sqrt{2}}  \left[ s_{2\beta} (\mu_1 - \mu_2) -c_{2\beta} \mu_3 \right] \,,  \nonumber \\
m_{\eta^0 \eta^0}^2 \equiv& m^2_{\eta^0} = m^2_\Delta + \left(\lambda_{\Delta 1} + \lambda_{\Delta 2} \right) v^2_\Delta + \frac{\bar\lambda_8 c^2_b +\bar\lambda_9  s^2_\beta}{2} v^2 + \frac{\bar\lambda_{12} s_{2\beta}}{2} v^2\,, \label{eq:CP-odd_M}
\end{align}
with $\bar\lambda_{8}=\lambda_8 + \lambda'_8$ and $\bar\lambda_{9}=\lambda_9 + \lambda'_9$. 
From Eq.~(\ref{eq:v_D}), it is known that $ c^2_\beta \mu_1+ s^2_\beta \mu_2 + c_\beta s_\beta \mu_3  \sim {\cal O}(v_\Delta)$.  Neglecting term of ${\cal O}(v_\Delta)$, one can see that $m^2_{G^0 G^0}$, $m^2_{G^0 A^0}$, and $m^2_{G^0 \eta^0}$ are negligibly small.  Thus, to a good approximation, $G^0$ represents the neutral Goldstone boson.  Moreover, if we further demand that the factor $s_{2\beta} (\mu_1 - \mu_2) - c_{2\beta} \mu_3$ in $m^2_{A^0 \eta^0}$ vanish, $A^0$ and $\eta^0$ decouple from each other.  To understand the correlations among $\mu_1$, $\mu_2$, and $\mu_3$ under the conditions:
\begin{align}
\begin{cases}
c^2_\beta \mu_1+ s^2_\beta \mu_2 + c_\beta s_\beta \mu_3&=\epsilon_\Delta \,, \\
s_{2\beta} (\mu_1 - \mu_2) - c_{2\beta} \mu_3&=0\,,
\end{cases}
\end{align}
where $\epsilon_\Delta$ is an parameter of ${\cal O}(v_{\Delta})$, we solve and obtain:
\begin{equation}
\mu_2 = \mu_1 t^{-2}_\beta + \epsilon_\Delta \left( 1- t^{-2}_\beta \right)
\,, ~ ~ 
\mu_3 = -2 t^{-1}_\beta (\mu_1 -\epsilon_\Delta )\,. \label{eq:mu_23}
\end{equation} 
For $\tan\beta=1$, we have $\mu_2 \simeq \mu_1$ and $\mu_3 \simeq -2\mu_1$.  For large $\tan\beta$, they can be approximated as:
\begin{equation}
\mu_2 \simeq \epsilon_\Delta + \mu_1 t^{-2}_\beta 
\,, ~~ 
\mu_3 \simeq -2 \mu_1 t^{-1}_\beta \,. 
\end{equation}
It is seen that the $\mu_{2,3}$ scale is determined by $\mu_1$ and $t_\beta$.

Similarly, the mass matrix for $G^+$, $H^+$ and $\delta^+$ is given by:
 \begin{equation}
  (G^- H^- \delta^- ) \left( \begin{array}{ccc}
    m^2_{G^- G^+} & m^2_{G^- H^+} & m^2_{G^- \delta^+} \\ 
    m^2_{G^- H^+} & m^2_{H^{-} H^{+}} & m^2_{H^- \delta^+} \\ 
    m^2_{G^- \delta^+} &  m^2_{H^- \delta^+} & m^2_{\delta^- \delta^+} \\ 
  \end{array} \right)  \left(\begin{array}{c}
    G^+  \\ 
    H^+  \\ 
    \delta^+ \\ 
  \end{array} \right)\,, \label{eq:mCH}
 \end{equation}
where the mass matrix elements
\begin{align}
m^2_{G^{-} G^+} & = \sqrt{2} v_\Delta \left( c^2_\beta \mu _1 + s^2_\beta \mu_2 + c_\beta s_\beta \mu_3 \right)- \frac{v^2_\Delta}{2}(c^2_\beta \lambda'_8 + s^2_\beta \lambda'_9 + \lambda'_{12} s_{2\beta})\,, \nonumber \\
m^2_{G^- H^+} & = -\frac{v_\Delta}{\sqrt{2}}  ( s_{2\beta}( \mu_1 - \mu_2) -c_{2\beta} \mu_3 ) + \frac{ \lambda'_8 - \lambda'_9 }{2} c_\beta s_\beta v^2_\Delta - \frac{\lambda'_{12} }{2} c_{2\beta} v^2_{\Delta} \,, \nonumber \\
m^2_{G^- \delta^+} & = -v \left(c^2_\beta \mu_1 + s^2_\beta \mu_2 + c_\beta s_\beta \mu_3 \right) + \frac{v_\Delta v}{2\sqrt{2}} \left(c^2_\beta \lambda'_8 + s^2_\beta \lambda'_9 + s_{2\beta} \lambda'_{12} \right)  \,,\nonumber \\
m^2_{H^- H^+} &\equiv m^2_{H^\pm} =  \frac{m^2_{12}}{c_\beta s_\beta} - \frac{\lambda_4 + \lambda_5}{2} v^2 - \frac{\lambda_6 v^2}{2} t^{-1}_\beta  - \frac{\lambda_7 v^2}{2} t_\beta \,,  \nonumber \\
& + \sqrt{2} v_\Delta \left( s^2_\beta \mu_1- c^2_\beta \mu_2\right) + \frac{ \mu_3 v_\Delta }{\sqrt{2} c_\beta s_\beta} \left( c^4_\beta + s^4_\beta \right) -\frac{v^2_\Delta}{2}\left( \lambda'_8 s^2_\beta + \lambda'_9 c^2_\beta\right)  \nonumber \\
& - \frac{v^2_\Delta}{2} \left( s_{2\beta} \lambda_{12} + (c^3_\beta + s^3_\beta) \bar\lambda_{12} \right)\,, \nonumber \\
m^2_{H^- \delta^+} &=  \frac{v}{2} \left( s_{2\beta}  (\mu_1 - \mu_2) - c_{2\beta} \mu_3 \right) - \frac{\lambda'_8-\lambda'_9}{2\sqrt{2}} c_\beta s_\beta v v_\Delta - \frac{\lambda'_{12}}{2\sqrt{2}} s^2_\beta v v_\Delta \,, \nonumber \\
m^2_{\delta^- \delta^+} &\equiv m^2_{\delta^\pm} = m^2_\Delta + \frac{2\lambda_9 + \lambda'_9 }{4} s^2_\beta v^2  + \frac{2\lambda_8  + \lambda'_8 }{4} c^2_\beta v^2 + \frac{2\lambda_{12} + \lambda'_{12}}{4} s_{2\beta} v^2\,.
\label{eq:Charged_M}
\end{align}
Analogous to the case of CP-odd scalar mass matrix, $m^2_{G^- G^+}$, $m^2_{G^- H^+}$, and $m^2_{G^- \delta^+}$  vanish if we drop the ${\cal O}(v_\Delta)$ terms.  Therefore, $G^\pm$ can be approximated as the charged Goldstone bosons.  If we further demand $ \mu_3=t_{2\beta} (\mu_1 - \mu_2)$, $m^2_{H^- \delta^+}$ also vanishes, and $H^\pm$ and $\delta^\pm$ decouple each other and are approximately the physical states.

Next, we discuss the CP-even scalars.  In 2HDMs, the physical $(H, h)$ states and the $(\Phi^0_1, \Phi^0_2)$ states are related by:
\begin{equation}
\begin{pmatrix}
H \\
h
\end{pmatrix}
= 
\begin{pmatrix}
  c_{\alpha} &  s_{\alpha} \\
 -s_{\alpha}  &  c_{\alpha}
\end{pmatrix}
\begin{pmatrix}
 \Phi^0_1 \\
 \Phi^0_2
\end{pmatrix}
\,,  \label{eq:H-Phi}
\end{equation}
where $\alpha$ is the mixing angle for CP-even scalars.  To derive the mass matrix in the $(h, H, \delta^0)$ states, it is more convenient to start from the basis of $(\Phi^0_1, \Phi^0_2)$.  Once the mass eigenvalues and eigenstates of $h$ and $H$ in the 2HDM are obtained, we then include $\delta^0$ to form a three-component basis $(h, H, \delta^0)$.  As we will explicitly see below, these will be approximately the physical states as long as ${\cal O}(v_\Delta)$ terms are ignored.

The $2\times 2$ mass matrix for $(\Phi^0_1, \Phi^0_2)$ is:
\begin{equation}
\frac{1}{2} 
\begin{pmatrix}
\Phi^0_1~ \Phi^0_2
\end{pmatrix}
\begin{pmatrix}
    m^2_{\phi_1 \phi_1} & m^2_{\phi_1 \phi_2} \\ 
    m^2_{\phi_1 \phi_2} & m^2_{\phi_2 \phi_2} 
\end{pmatrix}
\begin{pmatrix}
   \Phi^0_1  \\ 
   \Phi^0_2
\end{pmatrix}
\,, \label{eq:mPhi}
\end{equation}
where the matrix elements
\begin{align}
m^2_{\phi_1 \phi_1} 
&= \left(  m^2_{12} + \frac{v_\Delta \mu_3}{\sqrt{2}} \right) t_\beta + \lambda_1 v^2 c^2_\beta + \frac{3 \lambda_6 v^2}{2} c_\beta s_\beta - \frac{\lambda_7 v^2}{2}  s^2_\beta t_\beta \,, \nonumber \\
m^2_{\phi_1 \phi_2} 
&= -\left(  m^2_{12} + \frac{v_\Delta \mu_3}{\sqrt{2}} \right) + \lambda_{345} v^2 c_\beta s_\beta + \frac{3 v^2}{2} \left( c^2_\beta \lambda_6 + s^2_\beta \lambda_7 \right)\,, \nonumber \\
m^2_{\phi_2 \phi_2} 
&= \left(  m^2_{12} + \frac{v_\Delta \mu_3}{\sqrt{2}} \right) t^{-1}_\beta + \lambda_2 v^2 s^2_\beta - \frac{ \lambda_6 v^2}{2} c^2_\beta t^{-1}_\beta + \frac{3 \lambda_7 v^2}{2}  c_\beta s_\beta 
\,.  \label{eq:CP-even_M}
\end{align}
Using the parametrization in Eq.~\eqref{eq:H-Phi}, we obtain the eigenvalues of Eq.~\eqref{eq:H-Phi} as:
\begin{align}
m^2_{h,H} =  \frac{m^2_{\phi_1 \phi_1} + m^2_{\phi_2 \phi_2}}{2} \pm \frac{1}{2} \left[ \left( m^2_{\phi_1 \phi_1} - \phi^2_{\phi_2 \phi_2} \right)^2 + 4 (m^2_{\phi_1 \phi_2})^2 \right]^{1/2}\,,
\end{align}
and the mixing angle is determined by:
 \begin{equation}
 \tan{2\alpha} = \frac{2 m^2_{\phi_1 \phi_2}}{m^2_{\phi_1 \phi_1} -m^2_{\phi_2 \phi_2} }\,.
 \end{equation}

Using the rotational matrix, 
 \begin{equation}
 \begin{pmatrix} 
  -s_\alpha & c_\alpha  &0 \\
  c_\alpha & s_\alpha & 0 \\
  0 & 0 & 1 
  \end{pmatrix}  \,, 
 \end{equation}
we can transform the basis from $(\Phi^0_1, \Phi^0_2 , \delta^0)$ to $(h, H, \delta^0)$,  and the transformed mass matrix is:
\begin{equation}
\frac{1}{2} \left( h~ H~ \delta^0  \right)
\begin{pmatrix} 
m^2_{h} & 0 & m_{h \delta^0}^2 \\ 
0 & m_{H}^2 & m_{H \delta^0}^2 \\ 
m_{h \delta^0}^2 & m_{H \delta^0}^2 & m_{\delta^0 \delta^0}^2 
\end{pmatrix}
\begin{pmatrix} h \\ H \\ \delta^0 \end{pmatrix}\,,
\label{M_CP_even_2}
\end{equation}
where the additional elements
\begin{align}
m_{h \delta^0}^2  = & - \frac{v}{\sqrt{2}} \left[(\mu_1+\mu_2) s_{\beta-\alpha} -(\mu_1 - \mu_2) s_{\beta+\alpha} + \mu_3 c_{\beta+\alpha} \right] \nonumber \\
& -  \frac{v_\Delta v}{2}\left[ (\bar\lambda_8+\bar\lambda_9) s_{\beta-\alpha}+ (\bar\lambda_8 -\bar\lambda_9) s_{\beta+\alpha} -2 \bar\lambda_{12} c_{\beta+\alpha}\right]\,, \nonumber \\
m_{H \delta^0}^2  = &  -\frac{v}{\sqrt{2}} \left[ (\mu_1 + \mu_2) c_{\beta-\alpha} +(\mu_1-\mu_2) c_{\beta+\alpha} + \mu_3 s_{\beta+\alpha}\right] \nonumber \\
 & +\frac{v_\Delta v}{2} \left[ (\bar\lambda_8 + \bar\lambda_9)c_{\beta-\alpha} + (\bar\lambda_8 - \bar\lambda_9) c_{\beta+\alpha}+2 \bar\lambda_{12} s_{\beta+\alpha}\right]\,, \nonumber \\
 m_{\delta^0 \delta^0}^2  = & m^2_\Delta + 3 (\lambda_{\Delta 1} + \lambda_{\Delta 2}) v^2_\Delta + \frac{v^2}{2} \left( c^2_\beta \bar\lambda_8 + s^2_\beta \bar\lambda_9 + s_{2\beta} \bar\lambda_{12}\right)\,.
\end{align}
In the model, $m^2_{h\delta^0}$ and $m^2_{H\delta^0}$ are generally not small. Nevertheless, if we apply the conditions in Eq.~\eqref{eq:mu_23}, $m^2_{h\delta^0}$ and $m^2_{H\delta^0}$ can be rewritten as:
\begin{align}
m^2_{h \delta^0}  = &  - \sqrt{2} v \epsilon_\Delta  s_{\beta-\alpha} -  \frac{v_\Delta v}{2}\left[ (\bar\lambda_8+\bar\lambda_9) s_{\beta-\alpha}+ (\bar\lambda_8 -\bar\lambda_9) s_{\beta+\alpha} -\bar\lambda_{12} c_{\beta+\alpha}\right]\,, \nonumber \\
m_{H \delta^0}^2  = &  -\sqrt{2} v \epsilon_\Delta s_{\beta+\alpha}
  + \frac{v_\Delta v}{2} \left[ (\bar\lambda_8 + \bar\lambda_9)c_{\beta-\alpha} + (\bar\lambda_8 - \bar\lambda_9) c_{\beta+\alpha} + \bar\lambda_{12} s_{\beta+\alpha}\right]\,,
\end{align}
both being $\sim O(v_\Delta)$.  Comparing to the dominant diagonal elements $m^2_{h,H,\delta^0}$, the mixing effects among $h$, $H$, and $\delta^0$ are thus small.   Therefore, to the leading order in $v_\Delta$, we will neglect such mixing and $h$ and $H$ decouple from $\delta^0$.

{

\section{Scalar potential and trilinear couplings in the Higgs basis} \label{app:SP_HB}

The scalar potential in terms of the Higgs basis $H_{1,2}$ can be written as:
\begin{equation}
 V= V(H_1, H_2)+ V(\Delta) + V(H_1,H_2,\Delta)\,, 
 \end{equation}
where each term is more explicitly given by
\begin{align}
V(H_1,H_2) 
=& Y_1 H^\dagger_1 H_1 + Y_2 H^\dagger_2 H_2 +Y_3 ( H^\dagger_1 H_2 + \mbox{H.c.})
+ \frac{1}{2} \Lambda_1 ( H^\dagger_1 H_1)^2   \nonumber \\
 &+  \frac{1}{2} \Lambda_2 (H^\dagger_2 H_2)^2 + \Lambda_3  H^\dagger_1 H_1 H^\dagger_2 H_2 + \Lambda_4  H^\dagger_1 H_2 H^\dagger_2 H_1 +  \left[\frac{1}{2}\Lambda_5(H^\dagger_1 H_2)^2  \right. \nonumber \\
 &\left.+ \Lambda_6 (H^\dag_1 H_1) (H^\dag_{2} H_1) + \Lambda_7 (H^\dag_2 H_2) (H^\dag_2 H_1)
 + \mbox{H.c.} \right] \,,  \label{eq:v2_aH}
\\
V(\Delta )
=& M^2_\Delta Tr \Delta^\dagger \Delta + \Lambda_{\Delta 1} (Tr \Delta^\dagger \Delta)^2 + \Lambda_{\Delta 2} Tr (\Delta^\dagger \Delta)^2\,,  \label{eq:v2_bH}
\\
V(H_1,H_2,\Delta) 
=& \left( W_1 H^T_1 i\tau_2 \Delta^{\dagger}  H_1 + W_2 H^T_2 i \tau_2 \Delta^\dagger H_2 + W_3 H^T_1 i\tau_2 \Delta^\dagger H_2 + \mbox{H.c.} \right) \nonumber \\
&+ \left[ \Lambda_8 H^\dagger_1 H_1 + \Lambda_9 H^\dagger_2 H_2 + (\Lambda_{12} H^\dagger_1 H_2 + \mbox{H.c.}) \right] Tr \Delta^\dagger \Delta  \nonumber \\
& + \Lambda'_8  H^\dagger_1 \Delta \Delta^\dagger H_1 + \Lambda'_9 H^\dagger_2 \Delta \Delta^\dagger H_2 + (\Lambda'_{12}  H^\dagger_1 \Delta \Delta^\dagger H_2 + \mbox{H.c.})\,. \label{eq:v2_cH}
\end{align}
Using the representations of $H_{1,2}$ and $\Delta$ in Eq.~(\ref{eq:scalars}), the minimum conditions for the the VEVs of $H_{1,2}$ and $\Delta$ can be obtained as:
\begin{subequations}
\begin{align}
Y^2_1 + \frac{\Lambda^2_1}{2} v^2 &= \sqrt{2} W_1 v_D - \frac{\bar\Lambda_8}{2} v^2_D \,, \nonumber \\
Y^2_3 + \frac{\Lambda^2_6}{2} v^2 &= \frac{ W_3}{\sqrt{2}} v_D - \frac{\bar\Lambda_{12}}{2} v^2_D \,, \nonumber \\
M^2_\Delta + \frac{\bar\Lambda_8}{2} v^2 & = \frac{v^2}{\sqrt{2}} \frac{W_1}{v_\Delta} - (\Lambda_{\Delta 1} + \Lambda_{\Delta 2} ) v^2_\Delta\,,
 \end{align}
 \end{subequations}
with $\bar\Lambda_{8}=\Lambda_8 + \Lambda'_8$ and $\bar\Lambda_{12} = \Lambda_{12} + \Lambda'_{12}$.

 From Eqs.~\eqref{eq:v2_aH}--\eqref{eq:v2_cH}, the doubly-charged Higgs mass is obtained as:
  \begin{equation}
  m^2_{\delta^{\pm\pm}} = m^2_\Delta + \frac{\Lambda_8 v^2}{2} + \Lambda_{\Delta 1} v^2_\Delta\,. 
  \end{equation}
The mass splittings in the Higgs triplet are:
 \begin{align}
  m^2_{\delta^{\pm}}- m^2_{\delta^{\pm\pm}} & = \frac{\Lambda_{\Delta 2} v^2_\Delta}{2} + \frac{\Lambda'_8 v^2}{4}\,, \nonumber \\
   m^2_{\delta^{0}}- m^2_{\delta^{\pm\pm}} & =(2\Lambda_{\Delta 1} + 3 \Lambda_{\Delta 2} ) v^2_\Delta + \frac{\Lambda'_8 v^2}{2}\,.
 \end{align}
 The trilinear couplings of the neutral scalars to the charged Higgs scalars with $c_{\beta-\alpha}=0$ are given by:
  \begin{equation}
  {\cal L}_{H_i SS} = -v \left[ \lambda_{H_i \delta^{--} \delta^{++}} H_i  \delta^{--} \delta^{++} + \lambda_{H_i \delta^{-} \delta^{+}}  H_i \delta^- \delta^+  + \lambda_{H_i H^{-} H^{+}}  H_i H^-H^+ \right]\,,
  \end{equation}
 where the couplings:
  \begin{align}
  \lambda_{h \delta^{--} \delta^{++}} & = \Lambda_8 \,, ~ \Lambda_{H \delta^{--} \delta^{++}} = -\Lambda_{12} \,, \nonumber \\
  \lambda_{h \delta^- \delta^+} & = \Lambda_8 + \frac{\Lambda'_8}{2}\,, ~ \lambda_{H \delta^- \delta^+} = -\Lambda_{12} - \frac{\Lambda'_{12}}{2} \,, \nonumber \\
  \lambda_{h H^- H^+} & = \Lambda_3 \,, ~ \lambda_{H H^- H^+} = -  \Lambda_7\,.
  \end{align}

}

\end{document}